\documentclass[aps,twocolumn,preprintnumbers,superscriptaddress]{revtex4} 
\usepackage[normalem]{ulem}
\usepackage{amsmath}
\usepackage{enumerate}
\usepackage{amsfonts}
\usepackage{epsfig}
\usepackage{amsfonts}
\usepackage{dsfont}
\usepackage{ upgreek }

\usepackage{color}

\def\g{\gamma}

\def\Tr{\mathop{\rm Tr}}
\def\le{\left}
\def\ri{\right}

\newcommand\de{{\ensuremath{{\delta}}}}

\newcommand{\be}{\begin{equation}}
\newcommand{\ee}{\end{equation}}
\newcommand{\ben}{\begin{enumerate}}
\newcommand{\een}{\end{enumerate}}
\newcommand{\bea}{\begin{eqnarray}}
\newcommand{\eea}{\end{eqnarray}}

\newcommand\ket[1]{\ensuremath{\lvert{#1}\rangle}}
\newcommand\bra[1]{\ensuremath{\langle{#1}\rvert}}

\newcommand\Om{\Omega}

\begin{document}

\preprint{MIT-CTP/488§6}

\title{Canonical Universality}
                             
\author{Anatoly Dymarsky} \affiliation{Department of Physics and Astronomy, University of Kentucky, Lexington, KY 40506\\
Skolkovo Institute of Science and Technology, Skolkovo Innovation Center, Moscow, Russia 143026\\[1.5pt] }
\author{Hong Liu} \affiliation{Center for Theoretical Physics,
Massachusetts
Institute of Technology,
Cambridge, MA 02139}

\begin{abstract}
An isolated quantum system in a pure state may be perceived as thermal if only substantially small fraction of all degrees of freedom is probed. We propose that in a chaotic quantum many-body system all states with  sufficiently small energy fluctuations are approximately thermal. We refer to this hypothesis as Canonical Universality (CU). The CU hypothesis complements the Eigenstate Thermalization Hypothesis (ETH) which proposes that for chaotic systems individual energy eigenstates are thermal. Integrable and MBL systems do not  satisfy CU. We provide theoretical and numerical evidence supporting the CU hypothesis. 
\end{abstract}

\maketitle

Consider an isolated quantum system in a pure state  $\ket{\psi}$. We assume $\ket{\psi}$ belongs to a sufficiently narrow energy band
\bea
\label{state}
\ket{\psi}=\sum c_n \ket{E_n},\quad E_n\in [E-\Delta E,E+\Delta E]\ ,
\eea
where $\ket{E_n}$ are eigenstates of energies $E_n$. 
We probe the system with an operator $A$, which explores only a small fraction of all degrees of freedom.  For example $A$ could be acting on a substantially small subsystem of the full system. In this case Canonical Typicality \cite{Goldstein,Popescu} ensures that there is a high probability that the expectation value  $\bra{\psi}A \ket{\psi}$
would be approximately thermal  (microcanonical)
\bea
\label{vev}
\bra{\psi}A \ket{\psi}\simeq { \mathcal N}^{-1} \sum A_{nn}\equiv A^{\rm micro}\ ,
\eea
for a  typical i.e.~random state \eqref{state}. Corrections to \eqref{vev} are suppressed by ${\mathcal N}^{-1/2}$, where the number of energy levels inside the energy band $\mathcal N=\int\limits_{E-\Delta E}^{E+\Delta E} \Omega\,  dE$ is assumed to be exponentially large and $\Om (E)$ is the density of states. Although \eqref{vev} is true  for most states, there might be states inside the band that are not thermal in the sense of \eqref{vev}, as is normally the case for energy eigenstates of integrable models. 

In this paper we propose that for quantum chaotic systems {\it all} states  of the form \eqref{state} with a sufficiently small $\Delta E$
are thermal.
To investigate possible deviation of $\psi$ from thermal equilibrium (as measured by the operator $A$), 
we introduce functions $A^{\rm max}$ and $A^{\rm min}$ as the maximal (minimal) possible values of $\bra{\psi}A\ket{\psi}$ for all normalized states $\psi$ of the form \eqref{state},
\bea
&&A^{\rm max}(E,\Delta E)=\max_\psi\  \bra{\psi}A\ket{\psi}\ ,\\
&&A^{\rm min\,}(E,\Delta E)=\min_\psi\  \bra{\psi}A\ket{\psi}\ .
\label{min}
\eea
Assuming a discrete spectrum, $A^{\rm max/min}$ is simply the maximal (minimal) eigenvalue of a hermitian ${\mathcal N}\times {\mathcal N}$ matrix $A_{nm}$ with $n,m$ satisfying $E-\Delta E\leq E_n,E_m\leq E+\Delta E$. As such it is a monotonic function of $\Delta E$ for fixed $E$.  
The functions $A^{\rm max/min} - A^{\rm micro}$ specify the maximal/minimal possible deviation from thermal behavior, as measured by the operator $A$, for all states~\eqref{state}.
It is then convenient to introduce a function  $\Delta E(E,x)$ defined through \footnote{We will often suppress one of the arguments of $\Delta E(E,x)$, writing it simply as $\Delta E(x)$, whenever the implied value of $E$ is not ambiguous.} 
\bea
\label{xplus}
A^{\rm max}(E,\Delta E(x))-A^{\rm micro}(E)&=&x\ ,\quad {\ \rm for\ } x> 0\ ,\\
A^{\rm min}(E,\Delta E(x))-A^{\rm micro}(E)&=&x\ ,\quad  {\ \rm for\ } x< 0\ .
\label{xminus}
\eea
Function $\Delta E(x)$ specifies minimal width of an energy band that includes at least one non-thermal state that exceeds some ``tolerance level" $x$. Note that instead of $A^{\rm micro}$ we could  use another definition of thermal expectation value, say the canonical one. Normally we will consider $x$ to be much larger than the ambiguity associated with different ways to define a thermal expectation value.
It is convenient to normalize $A$ rendering it dimensionless. In case of a finite-dimensional local Hilbert space we require $\parallel\hspace{-4pt}A\hspace{-4pt}\parallel=1$, which limits  $|x|\leq 1$.

The operator $A$ could be a macro-observable associated with some extensive quantity. Qualitatively, in this case $\Delta E(x)$ specifies the minimal amount of energy  fluctuations necessary to deviate from macroscopic thermal equilibrium (MATE), as defined in \cite{GoldsteinTE1,GoldsteinTE2}. For operators $A$ that are confined to a particular small subsystem, one can speak of energy fluctuations necessary to deviate from the microscopic thermal equilibrium (MITE). In the latter case $\Delta E(x)$ can be defined without specifying any particular $A$. Rather,  for a system in a state $\psi$ we define the reduced density matrix of the subsystem $\rho^\psi$, and introduce $x$ via the trace distance or other appropriate norm, 
\bea
\label{densitymatrix}
x=\max_\psi ||\rho^\psi -\rho^{\rm micro}||\ .
\eea
The maximum here is taken over all states \eqref{state} belonging to the band of width $\Delta E=\Delta E(x)$,  and $\rho^{\rm micro}(E)$ is the reduced density matrix of the microcanonical ensemble.

We propose that in a chaotic system, for  any operator $A$, up to exponentially small corrections $\Delta E (x)$ can be described by a smooth function  $\g(x)$, modulo a possible non-analyticity at $x=0$, 
\be \label{uro}
\Delta E (E, x) = \g(E, x) + O\le(\Om^{-1}\ri)  \ .
\ee 
From the definition of $\Delta E(x)$,  $\g(x)$ should be a  monotonically non-decreasing function for $x>0$ (non-increasing for $x<0)$. In the thermodynamic limit $V\rightarrow \infty$ with $E/V$ kept fixed,
for small $x$, $\g(x)= \g_0 x^\delta + \cdots$.  Leading exponent $\de$ depends on operator, for a generic one $\delta=2$. Coefficient $\gamma_0$ may be volume-dependent, but is not smaller than an inverse power of a characteristic system size, $\gamma_0\geq L^{-a}$, for some $A$-dependent $a\geq 0$. 

\begin{figure}
\includegraphics[width=.45\textwidth]{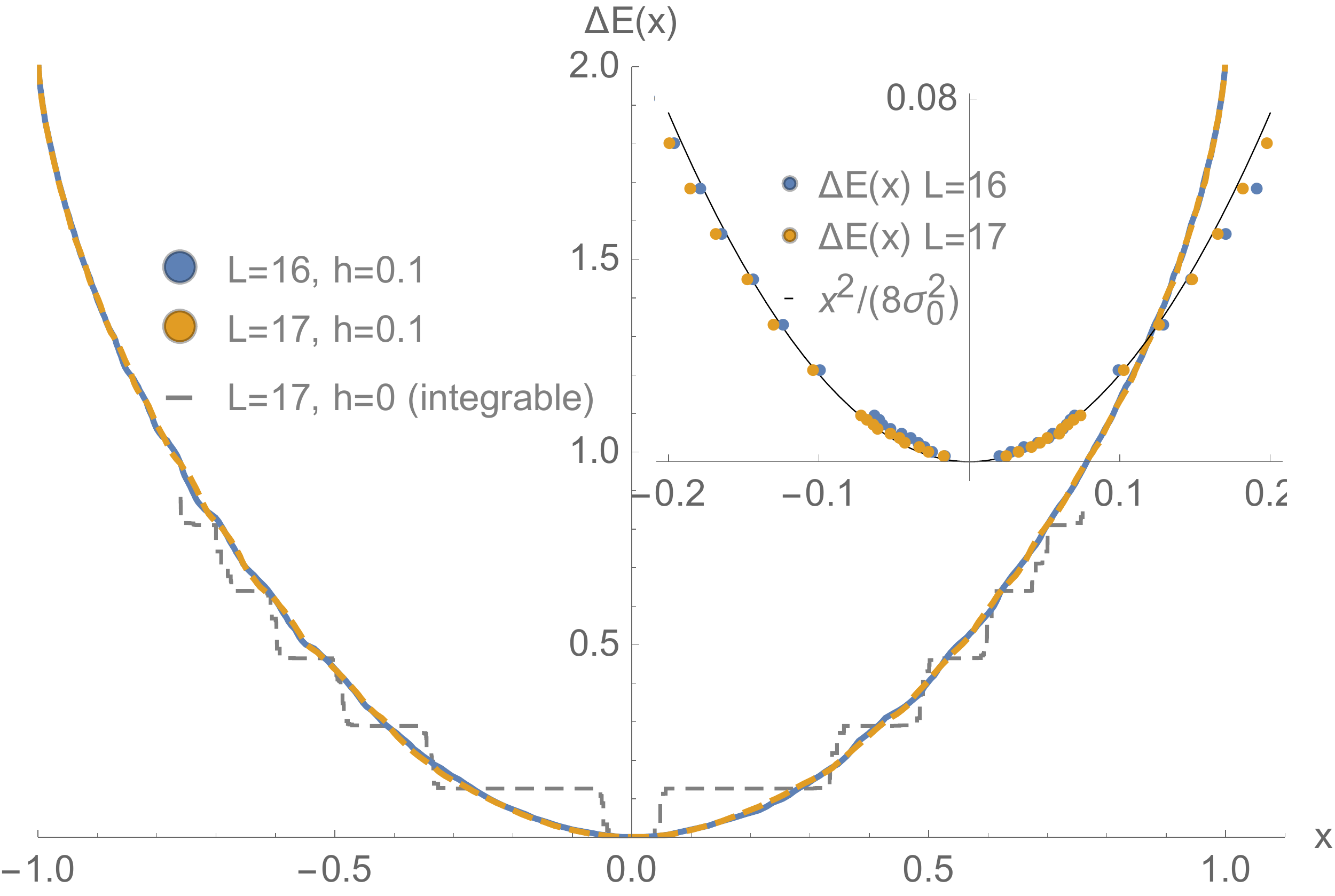}
\caption{
Numerical plot of $\Delta E(E=0,x)$ for the operator \eqref{AO} and $L=16,17$ in the non-integrable case $h=0.1$ superimposed with the integrable case $L=17, h=0$ (dashed line). While $\Delta E(x)$ in the non-integrable case is smooth for all $x$, in the integrable case it exhibits a characteristic plateau behavior. In the limit $L\rightarrow \infty$, the plateau $\Delta E\approx 0$ will stretch to at least $|x|\simeq 0.06$.  Inset: zoomed region of small $x$. Numerical values for $\Delta E(x)$ and $L=16,17$ and $h=0.1$ superimposed with the theoretical fit $\Delta E(x)=x^2/(8\sigma_0^2)$ (black line).}
\label{fig:O}
\end{figure}

\begin{figure}[b]
\includegraphics[width=.45\textwidth]{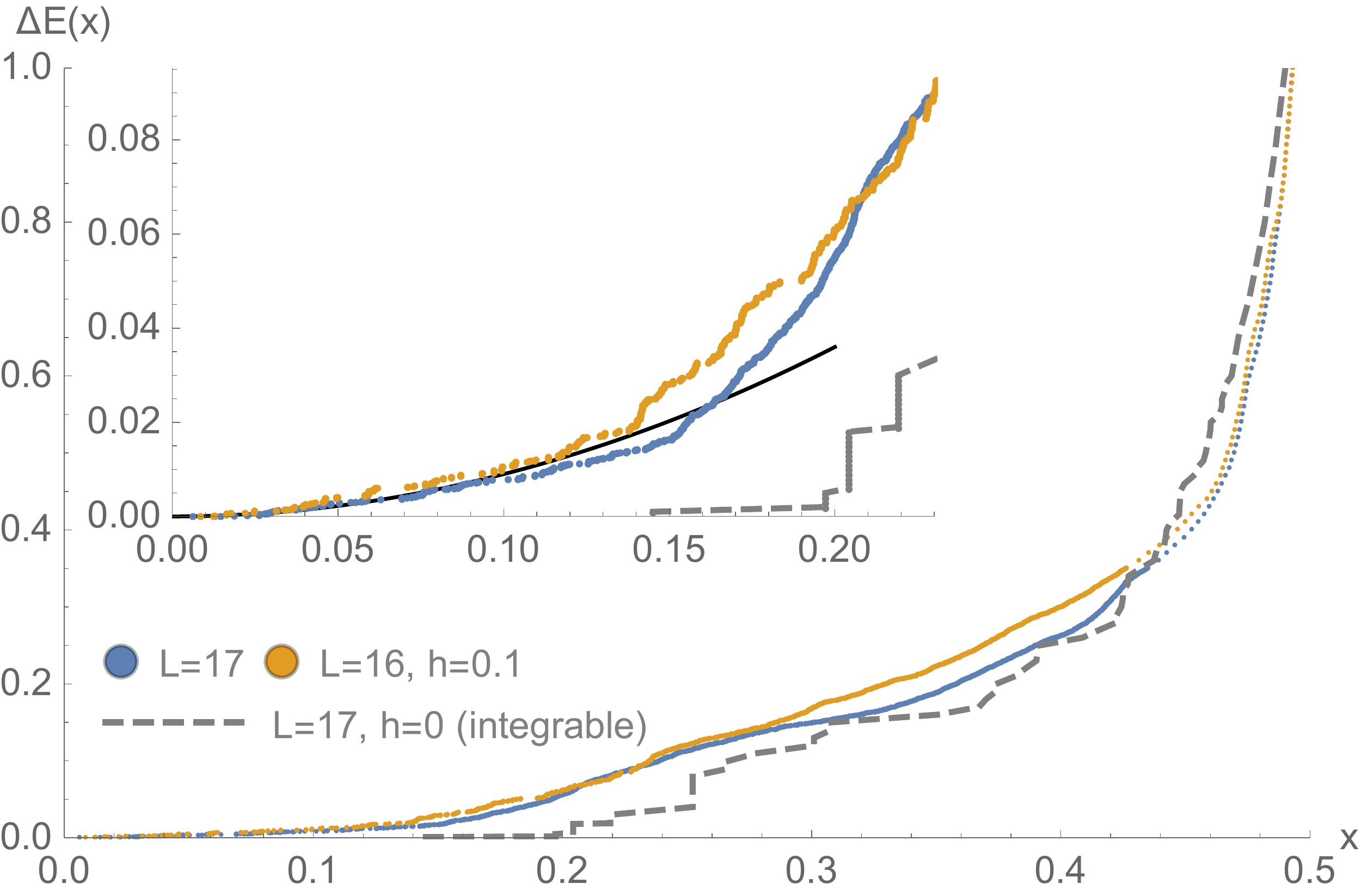}
\caption{Numerical plot of $\Delta E(E=0,x)$  \eqref{ydefmain} for the subsystem consisting of one leftmost spin $i=1$.  Data points for $L=16,17$ for non-integrable $h=0.1$ case superimposed with the integrable model $L=17, h=0$ results (dashed line). Inset: zoomed region of small $x$ superimposed with $x^2$ fit.}
\label{fig:E}
\end{figure}

For large but finite systems $\g(x)$ remains strictly positive for $x\neq 0$ and is zero only at $x=0$.  Thus to deviate from thermal equilibrium by a small amount $x$, one has to consider states built from energy eigenstates spanning a sufficiently wide interval $\Delta E=\g(x) > 0$. In particular, for the value of $x$ below the accuracy of a measurement, all states in an energy band $\Delta E < \g(x)$ are thermal. We will refer to~\eqref{uro} and properties of $\g(x)$ as Canonical Universality (CU). While Canonical Typicality establishes that {\it typical} states from a narrow energy band are approximately thermal with an exponential precision, Canonical Universality postulates that {\it all} states from a sufficiently narrow band  \eqref{state} are approximately thermal with the precision controlled by the band size $\Delta E$ \footnote{Both, the typicality arguments of \cite{Goldstein,Popescu} and universality, proposed in this paper, compare $\bra{\psi}A\ket{\psi}$ with the expectation of $A$ in the microcanonical ensemble. The relation between the latter and the expectation in the canonical ensemble is a secondary issue. Hence, more accurately we should call our conjecture ``microcanonical universality". Nevertheless following the terminology established  in \cite{Goldstein}, we use the language of canonical universality.}. Clearly~\eqref{uro} is {\it not} satisfied in integrable or  MBL systems, where  expectation values in neighboring energy levels could differ by a finite amount, i.e. $\Delta E(x)$ can develop characteristic plateau $\Delta E\sim \Omega^{-1}$ for a finite range of $x$ (as we will later see in Fig.~\ref{fig:O} and Fig.~\ref{fig:E}).
In particular,  this means the behavior of $\Delta E(x)$ can be used as an order parameter  to distinguish chaotic and non-chaotic phases.

Smooth behavior of $\g(x)$ requires that for $\Delta E \sim \Omega^{-1}$, $x$ should be exponentially small. In other words, if we consider nearby states $E_m$ and $E_n$, matrix elements $A_{mm}$ and $A_{nn}$ must be exponentially close \footnote{This is the requirement of strong ETH that all energy eigenstates are thermal i.e.~there are no ``outliers" \cite{SETH}.} and $A_{mn}$ must be exponentially small.  
This is reminiscent of the Eigenstate Thermalization Hypothesis (ETH) \cite{Deutsch,Srednicki}, which proposes that matrix elements $A_{nm}$ in energy eigenbasis  have a form~\cite{Srednicki1999}
\bea
\label{eth}
A_{nm}=A^{\rm eth}(E)\delta_{nm}+\Omega^{-1/2}(E) f(E,\omega)  r_{nm}\ , \\
E=(E_n+E_m)/2\ ,\quad \omega=(E_m-E_n)\ . \nonumber
\eea
Here $A^{\rm eth}$ and $f$ are smooth function of their arguments, and  ``fluctuations" $r_{nm}$ by definition have unit variance. 
CU~\eqref{uro}  indirectly constrains~\eqref{eth} when there is an exponentially large number of states between $n$ and $m$.

If we assume $r_{mn}$ are independently distributed, compatibility of~\eqref{uro} and~\eqref{eth} will become apparent. 
It is convenient to replace $A^{\rm micro}$ of~(\ref{xplus}) and (\ref{xminus}) by
$A^{\rm eth}(E)$, and similarly $\rho^{\rm micro}$ of \eqref{densitymatrix} by the universal density matrix of the subsystem ETH introduced in \cite{Dymarskyetal}. From the results for a band random matrix~\cite{Pastur} one finds that 
$\g(x)$ can be expressed in terms of $f$. To illustrate this relation we first consider a special case, taking variance $\sigma^2=|f(E,\omega)|^2$ of the off-diagonal matrix elements to be constant for $|\omega|\leq 2\Delta E$, and $r_{mn}$  to  
be a Gaussian Random Matrix compatible with the global symmetries of the problem \footnote{This choice is suggested by both, theoretical expectations that matrix elements in a narrow shell are well represented by a Gaussian Ensemble \cite{Review}, as well as numerical studies confirming Gaussian form of the distribution of $r_{nm}$ for various non-integrable models \cite{diagonal,offdiagonal,Dymarskyetal}. New numerical evidence supporting GOE form of $r_{nm}$ is provided in the supplementary materials.}. 
When $\Delta E$ is sufficiently small so that the total number of energy levels inside the band~\eqref{state} can be approximated as $\mathcal N\approx 2\Omega(E) \Delta E$, value of $x$ from~(\ref{xplus},\ref{xminus}) is readily given by the largest (smallest) eigenvalue of the Gaussian Random Matrix $R_{nm}=\Omega^{-1/2}f\,r_{nm}$ of size $\mathcal N$, 
\bea
x=2\sqrt{\mathcal N}\Omega^{-1/2}\sigma \; \Rightarrow \; \Delta E(x)={x^2\over 8 \sigma^2} \ .
\label{curmm}
\eea
Relaxing that $\sigma=f$ or $\Om$ are constant within the energy band  will result in higher power corrections in $x$ \footnote{For example, assuming constant $f$ and non-zero $T^{-1}=\partial \log \Om /\partial E$ one can calculate next order correction to be  
$\Delta E(x)={x^2\over 8 \sigma^2}-{x^6\over 3072 \sigma^6 T^2}+\dots $, with higher order terms being non-universal.}.

In full generality, band random matrix approximation provides the following bound on $x^2$
(see supplementary materials),
\bea
\label{inequality}
x^2(\Delta E)\leq 8\int\limits_0^{2\Delta E} |f(E,\omega)|^2 d\omega\ .
\eea 
The behavior of the right hand side of~\eqref{inequality} can be deduced from the connected
two-point function $C(t)=\langle E| A(t) A(0)|E\rangle_c$  associated with energy $E$~\cite{FDT,Review},
\bea
\label{intT}
\int\limits_0^\infty dt\, {\sin(t\Delta E)\over t \pi} \,{\rm Re}\, C(t) = \int\limits_0^{\Delta E} |f(E,\omega)|^2 d\omega\ .
\eea
Because of oscillatory behavior of ${\sin(t\Delta E)/t}$  integral in the left hand side of \eqref{intT} can be approximated as an averaged value of $C(t)$ on an interval $0\leq t<T\sim \Delta E^{-1}$. For a local operator $A$ and a translationally invariant system, let us consider thermodynamic limit $L\rightarrow \infty$, while always keeping $\Delta E^{-1}$ smaller than thermalization time $\tau$, time when  $C(t)$ becomes $L$-dependent. In case of a diffusive quantity $A$ this is Thouless time $\tau \sim L^2$.

Behavior of $f(\omega)$ for $|\omega|\lesssim \tau^{-1}$ is expected to be volume-dependent \cite{Review}, but remarkably \eqref{intT} shows that the integral of $|f(\omega)|^2$ for  $\Delta E\gtrsim \tau^{-1}$ only depends on universal ($L$-independent) behavior of $C(t)$.
After taking thermodynamic limit $C(t)$ is expected to vanish as $t\rightarrow \infty$.
Thus the integral in \eqref{intT} will go to zero when $\Delta  E\rightarrow 0$. This means $x^2(\Delta E)\rightarrow 0$ when $\Delta E\rightarrow 0$, implying
$\gamma(x)$ for $x\neq 0$ should remain strictly positive even after taking thermodynamic limit. 

This conclusion is perhaps too strong as it is based on an unjustified  assumption that $r_{nm}$ are independently distributed. Still this is expected to hold for bands not exceeding Thouless energy $\Delta E\lesssim \tau^{-1}$. Say, for a diffusive system in one dimensions $C(t)\sim t^{-1/2}$ and \eqref{inequality} gives $x^2(\Delta E\sim\tau^{-1})\lesssim L^{-1}$. Assuming $\Delta E(x)$ for $\Delta E\geq \tau^{-1}$ is of the form $\Delta E(x)\approx \gamma_0 x^\delta$,  we readily find $\gamma_0\geq L^{\delta/2-2}$ \footnote{In case of spatial disorder, when the transport  becomes subdiffusive  $C(t)\sim t^{-\gamma}$ \cite{gamma},  one finds $\gamma_0\geq L^{\delta/2-1/\gamma}$.}.

Finally we note that operators of the type $A = i [H, B]$ for some $B$, to which we will refer as descendant operators,
exhibit the behavior $\g(x) \propto x^\delta$ with $\delta<2$.
 As we discuss in supplementary materials descendant operators must satisfy the inequality $\Delta E(x)\ge |x|/(2\hspace{-4pt}\parallel\hspace{-4pt}B\hspace{-4pt}\parallel)$. Thus for such operators $\Delta E (x)$ at $x\rightarrow 0$ increases much faster than the generic $x^2$ behavior.
This is physically sensible as such operators are exactly thermal in an energy eigenstate and to deviate from thermal equilibrium one would need a larger amount of energy fluctuations.

We illustrate different behavior $\Delta E(x)$ with help of Ising  spin-chain model with the Hamiltonian
\be
\label{1dspinchain}
H=-\sum_{i=1}^{{L}-1}  \sigma_z^i\otimes \sigma_z^{i+1}+g\sum_{i=1}^{{L}} \sigma^i_x+h\sum_{i=1}^{{L}} \sigma^i_z\ .
\ee 
For comparison we  present the results for a non-integrable $g=1.05, h=0.1$ and integrable $g=1.05, h=0$ cases. 
{For a given energy band, the value of $x(E,\Delta E)$ can not be smaller than the variations of the thermal expectation value $A^{\rm micro}(E')$ for $E'$ inside the interval $|E-E'|\leq \Delta E$. In the thermodynamic limit these variations will be suppressed as $1/V$, which provides an upper bound on the convergence rate of $\Delta E(E,x)$. 
To minimize effects associated with finite $L$, we present numerical results for the local operator
\be
\label{AO}
A={g \sigma_z^1-{\rm h}\sigma_x^1\over \sqrt{g^2+{\rm h}^2}}\ ,\quad  {\rm h}=0.1\ ,
\ee
which has very small variance of $A^{\rm micro}(E)$ in a wide range around $E=0$.} We choose center of the band to be at $E=0$ at it corresponds to maximal density of states and infinite temperature. 
The plot shown in Fig.~\ref{fig:O}, supports the conclusion that $\Delta E(x)$ in the non-integrable case quickly becomes a smooth function, which, for small $x$, is well approximated by \eqref{curmm}. The plot for the same operator \eqref{AO} in the integrable case $h=0$, also shown in Fig.~\ref{fig:O}, clearly indicates $\Delta E(x)$ remains non-smooth and exhibits a characteristic plateau at small $\Delta E$.
\begin{figure}
\includegraphics[width=.45\textwidth]{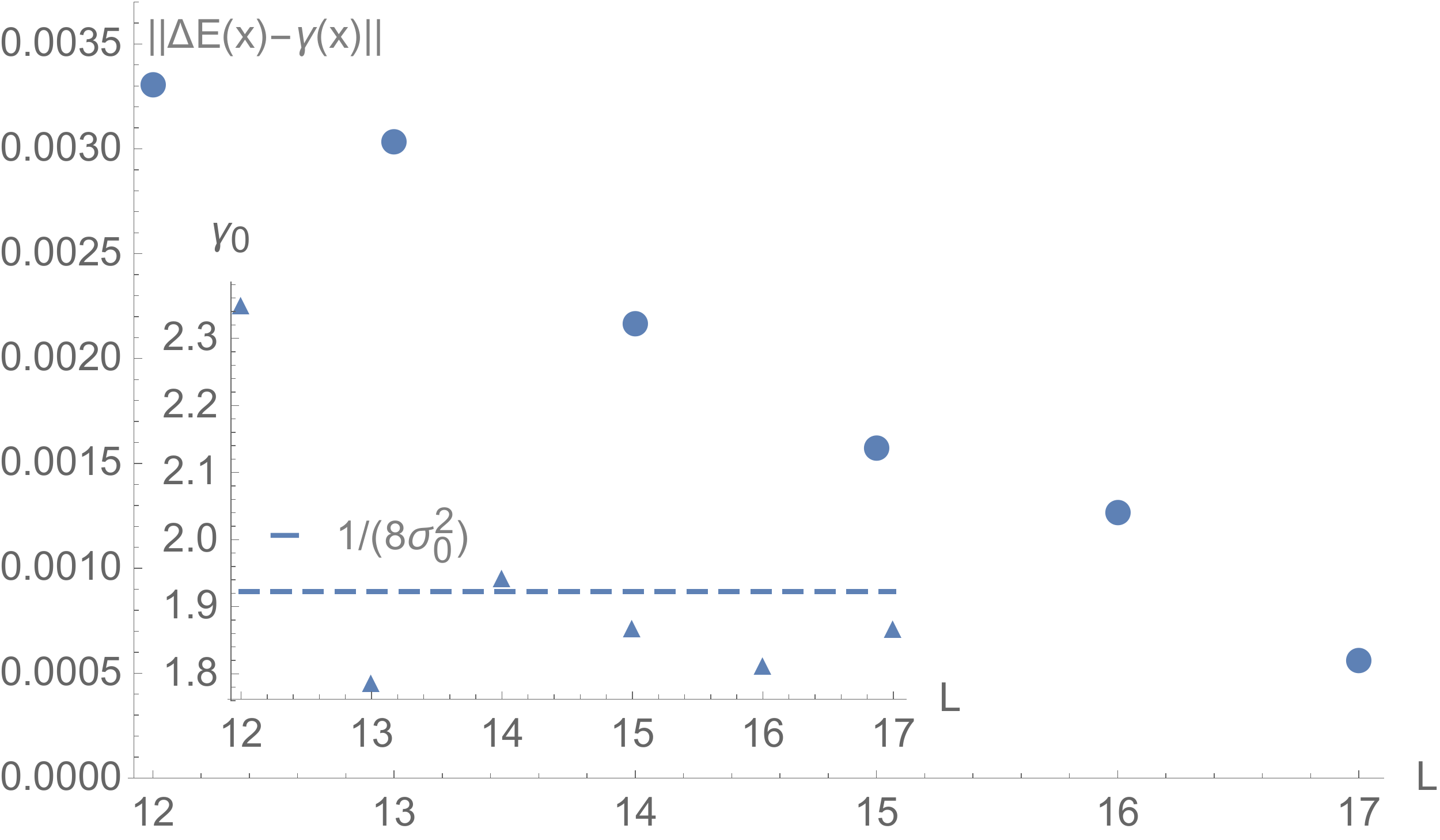}
\caption{Plot of $||\Delta E(x)-\g(x)||$  for the operator \eqref{AO},  $h=0.1$, and different $L=12-17$. Inset: plot of $\g_0$ for the best fourth-order polynomial fit (for the region $\Delta E\leq 0.2$) superimposed with the constant value $1/(8\sigma_0)^2$ (dashed horizontal line).}
\label{fig:norm}
\end{figure}

To access convergence of $\Delta E(x)$ to a smooth $\g(x)$, we introduce the ``deviation  norm" $||\Delta E(x)-\g(x)||$, defined as the variance of the difference $\Delta E_i -\g(x_i)$ for $x_i=x(\Delta E_i)$. The intervals $\Delta E_i$ represent incremental increase of the number of levels inside the interval, ${\mathcal N}(\Delta E_{i+1})={\mathcal N}(\Delta E_{i})+1$, and $\g(x)$ is a best degree-four  polynomial fit of $\Delta E(x)$. The plot in Fig.~\ref{fig:norm} shows a rapid decrease of the deviation norm with the system size, supporting \eqref{uro}. Numerical values of $\g_0$, which we define as the $x^2$ coefficient of the best polynomial fit, are shown  for different $L$ in the inset of Fig.~\ref{fig:norm}. The results are consistent with the proposal that $\g_0$ may at most polynomially depend on $L^{-1}$.

\begin{figure}
\includegraphics[width=.45\textwidth]{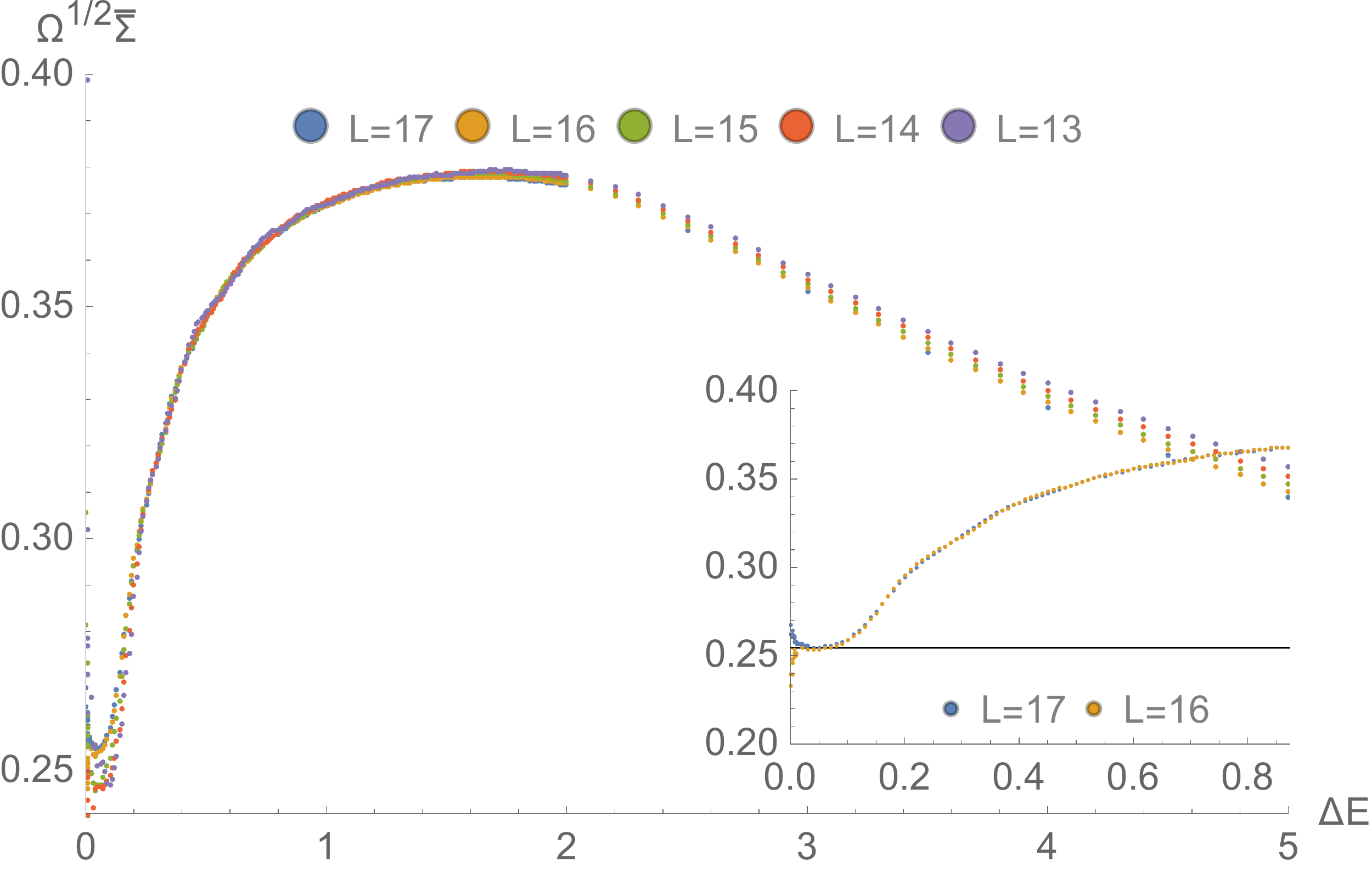}
\caption{Plot of $\Omega^{1/2}(0)\bar\Sigma(0,\Delta E)$ for operator \eqref{AO}, $h=0.1$ and different $L=13-17$. Inset: plots for $L=16,17$ at small $\Delta E$ superimposed with the constant value $\sigma_0=0.255$ (horizontal line).}
\label{fig:SO}
\end{figure}

As we discussed earlier, if $r_{nm}$ are random and independent and $f$ is approximately constant at small $\omega$, $\delta=2$ and $\g_0=1/(8|f|^2)$. We now test this relation numerically. For this purpose it is convenient to introduce running average variance 
\bea
\bar{\Sigma}^2(E,\Delta E)={1\over {\mathcal N}(\mathcal N-1)}\sum_{n\neq m} |A_{nm}|^2\ ,
\eea
where the sum is over all states inside the band $[E-\Delta E, E+\Delta E]$. In the thermodynamic limit, when \eqref{eth} applies and for sufficiently narrow $\Delta E\gg \Omega^{-1}$, such that $\Omega(E)$ is approximated constant within the energy band, 
\bea
\label{bs2}
{\Omega(E)} \bar\Sigma^2(E,\Delta E)=\int_{-1}^{1} dt\, (1-|t|)\left|f(E,2\Delta E t)\right|^2 .\ \ \ \
\eea
The plot of $\Omega^{1/2}\bar\Sigma(\Delta E)$ for operator \eqref{AO}, $E=0$, and different $L$,  depicted in Fig.~\ref{fig:SO}, shows that for $\omega$ of order one $f(0,\omega)$ quickly approaches a universal $L$-independent form . The same conclusion is corroborated by the analysis of two-point function $\langle A(t) A(0)\rangle_c$ (see supplementary materials).

The inset of Fig.~\ref{fig:SO}  suggests that $\Omega(0)\bar\Sigma(0,\Delta E)$ at $\Delta E\rightarrow 0$, and hence $f(0,\omega \rightarrow 0)$, approach a constant $f= \sigma_0\approx 0.255$. This numerical value together with \eqref{curmm} provide a good approximation for actual $\Delta E(x)$, as shown in the  inset of Fig.~\ref{fig:O}. 
Besides, $1/(8\sigma^2_0)$ and the value of $\g_0$ we read from the best polynomial fit of $\Delta E(x)$ are also reasonably consistent,  see the inset of Fig.~\ref{fig:norm}. This supports the assumption that $r_{mn}$ inside a substantially wide energy interval  are independently distributed.

Next, we discuss CU in the context of a subsystem, when the deviation from thermal equilibrium $x$ is defined through \eqref{densitymatrix}. 
In practice it is more convenient to define $x$ in terms of the Frobenius norm (for a one-spin subsystem considered below these definitions coincide), 
\bea
\label{Frobenius}
x^2=\Tr(\rho^\psi-\rho^{\rm micro})^2/2=(e^{-s_2}-e^{-s_0})/2\ .
\eea
Here $\rho^\psi$ is the reduced density matrix of the subsystem, and   
$\rho^{\rm micro}$  is the thermal density matrix, which in case of infinite temperature is  given by $\mathbb{I}/d$ ($d$ stand for the dimension of the Hilbert space of the subsystem). We have also introduced $s_2$ as the second Renyi entropy associated with the state  $\rho^{\psi}$, while $s_0\equiv \log(d)$. The definition  \eqref{Frobenius} emphasizes the role of entanglement entropy as a measure of proximity of the reduced state to the thermal one. Thermal behavior is associated with the maximal volume-law entanglement $s_2=s_0$ and $x=0$. This is in contrast to ``non-thermal'' energy eigenstates of integrable and MBL systems, which  exhibit sub-volume entanglement.

With help of the results of \cite{Convexball}, the problem of calculating $\Delta E(x)$ defined through \eqref{Frobenius} can be reformulated as a maximization problem on a unit sphere $\mathbb S^{d(d-1)}$,
\bea
\label{ydefmain}
x(\Delta E)={\max_{|\vec{c}|=1} \lambda_{\rm max} (\vec{c}\cdot \vec{{ \upvarsigma}})/\sqrt{2d}}\  .
\eea
Here $\lambda_{\rm max}$ denotes largest eigenvalue of a Hermitian matrix and $\upvarsigma^k$, for  $k=1\dots d(d-1)$, is the restriction of the full set of operators acting on the subsystem onto the energy band $[E-\Delta E, E+\Delta E]$.

For the subsystem consisting of one leftmost spin, $d=2$, and maximization in \eqref{ydefmain} can be readily performed. 
Numerical results for nonintegrable  and integrable cases are shown in Fig.~\ref{fig:E}. In the non-integrable case $\Delta E(x)$ is  smooth  and  is characterized by $\propto x^2$ behavior at small $x$. For the integrable case $\Delta E(x)$ is not smooth and exhibits a characteristic plateau near $\Delta E=0$.

Finally, we discuss the case of averaged quantities, e.g.~full magnetization $A_{x}=\sum_{i=1}^L \sigma^i_{x}/L$. First, we note that deviation of an averaged quantity from equilibrium requires deviation of all corresponding local quantities as well.
Accordingly $\Delta E(x)$ for an averaged quantity can not be smaller than $\Delta E(x)$ for the corresponding local operator, which ensures universality, i.e.~$\gamma(x)> 0$ for $x\neq 0$ for large but finite systems. Numerical plots show that $\Delta E(x)$ is smooth in the chaotic case, while non-smooth when the system is integrable. Volume dependence of $\Delta E(x)$ is more complicated. Analytic argument presented in the supplementary materials show that $\Delta E(x)$ must exhibit different scaling with $L$ for different values of $x$. 

Let us summarize our findings. We argued that for quantum chaotic systems, {\it all} states from a sufficiently narrow energy band must be approximately thermal in terms of microscopic and macroscopic equilibrium. This behavior, which we refer to as Canonical Universality, can be quantified in terms of function $\Delta E(x)$ that specifies maximal deviation from thermal equilibrium for states from a narrow energy band.  We propose that in the chaotic case for a general operator $\Delta E(x)= \g(x)$ becomes smooth and exhibits $\gamma \approx \gamma_0 x^\delta,\ \delta\ge 2$, behavior at small $x$.  We provided analytic and numerical evidence that $\g_0$ is at most polynomially dependent on $L^{-1}$. We expect that $\gamma_0$, which has dimension of energy, is related to the characteristic time-scale of thermalization $t_0$. Finally, we note that $\Delta E (x)$ provides an efficient way to distinguish chaotic systems from non-chaotic ones. In particular,  $\gamma_0$ can be used as an order parameter for transitions to chaos from integrable or MBL phases, providing new tools for these outstanding issues.\\[15pt]

We thank D.~Huse, J.~Lebowitz and A.~Polkovnikov for helpful comments and discussions. We would like to thank the University of Kentucky Center for Computational Sciences for computing time on the Lipscomb High Performance Computing Cluster.

\clearpage
\onecolumngrid

\centerline{\large \bf Supplementary Materials: Canonical Universality}
\vspace{.2cm}
\centerline{\large Anatoly Dymarsky${}^1$ and Hong Liu${}^2$} 
\vspace{.2cm}
\centerline{${}^1$
Department of Physics and Astronomy, University of Kentucky, Lexington, KY 40506}
\vspace{1pt}
\centerline{
Skolkovo Institute of Science and Technology, Skolkovo Innovation Center, Moscow, Russia 143026 }
\vspace{1.5pt}
\centerline{ 
${}^2$ Center for Theoretical Physics,
Massachusetts
Institute of Technology,
Cambridge, MA 02139}
\vspace{.5cm}

\twocolumngrid

\section{Band Random Matrices}
Let us consider an energy band of width $2\Delta E$ centered around $E$. We will keep $\Delta E$ and $E/V$ fixed, while volume grows $V\rightarrow \infty$. Assuming the system admits thermodynamic limit, both $A^{\rm eth}(E)$ and $f(E,\omega)$ introduced in \eqref{eth} are expected to smoothly depend on $E$ only through temperature $T=T(E/V)$. Hence one can approximate $(E_n+E_m)/2$ by the median energy of the band $E$. With an additional assumption that $\Delta E$ is narrow enough such that the density of states $\Omega$ within the band is approximately constant, \eqref{eth} can be rewritten as follows, 
\bea
\label{bm}
A_{nm}-A^{\rm eth}(E)\delta_{nm}=  {\sqrt{2\Delta E} \over  {\mathcal N}^{1/2}}v\left({n-m\over \mathcal N}\right) r_{nm}\ ,\  \\
v^2(t)=|f^2(E,2 t \Delta E)|\ , \ \ |t|\leq 1\ ,\ \ {\mathcal N}=2\Omega \Delta E\ .  \label{vdef} \nonumber
\eea
Assuming independent nature  (but not necessarily Gaussian form) of random variables $r_{nm}$, equation \eqref{bm} defines a band random matrix $\tilde A_{nm}=(A_{nm}-A^{\rm eth}(E))/\sqrt{2\Delta E}$, 
which was studies by Molchanov, Pastur, and Khorunzhii in  \cite{Pastur}. Namely, they consider a band random matrix $\tilde{A}_{nm}$, $n,m=1,\dots, \mathcal N$, with all elements being independently distributed, and the variance specified by an even non-negative function $v^2(t)$, 
\bea
\langle|\tilde{A}_{nm}|^2\rangle={\mathcal N}^{-1} v^2\left({n-m\over \mathcal N}\right)\ .
\eea
Under some technical assumptions, the generating function 
\bea
r(z)={1\over \mathcal N}\Tr{1\over z-\tilde{A}}\ ,
\eea
can be expressed in terms of an auxiliary function $r(z)=\int_{-1/2}^{1/2} r(t,z) dt$,
\bea
\label{rexp}
r(t,z)=-\sum_{i=0}^\infty {a_i(t)\over z^{2i+1}}\ ,\quad a_0(t)=1\ ,
\eea
while the latter satisfies a particular integral equation. This integral equation  can be rewritten as a system of recursive relations for $a_k(t)$,
\bea
\label{it}
a_{k+1}(t)=\sum_{p=0}^k a_p(t)\int_{-1/2}^{1/2}v^2(t-t')a_{k-p}(t')dt'\ .
\eea 
To obtain a bound on $a_k(t)$ we introduce $||a_k||=\max_{t\in[-1/2,1/2]}|a_k(t)|$ and immediately find 
\bea
\left|\left|a_{k+1}\right|\right| \leq \sum_{p=0}^k \left|\left|a_{p}\right|\right| \left|\left|a_{k-p}\right|\right| \int_{-1}^1 v^2(t) dt\ . 
\eea
These inequalities are saturated when $v^2(t)$ is a constant. In the latter case  the full nonperturbative solution for $r(t,z)$ is known, yielding 
\bea
\left|\left|a_{k}\right|\right| \leq 4^k{\Gamma(k+1/2)\over (k+1)! \sqrt{\pi}} \left(\int_{-1}^1 v^2(t) dt\right)^k\ .
\eea
Since $\int_{-1/2}^{1/2}a_k(t) dt\leq \left|\left|a_{k}\right|\right|$ we find that the expansion  \eqref{rexp} is convergent for 
\bea
z^2 \leq 4 \int_{-1}^1 v^2(t) dt\ .
\eea
Using the definition of $v^2$ \eqref{bm} and $\tilde{A}_{nm}$ we obtain \eqref{inequality}.

When $v^2(t)$ is approximately constant $v^2(t)=v_0^2+\delta v^2(t)$, \eqref{it}  can be solved perturbatively,  expanding in powers of $\delta v^2$, 
\bea
\nonumber
&&\int_{-1/2}^{1/2}a_k(t)dt= 4^k{\Gamma(k+1/2)\over (k+1)! \sqrt{\pi}}\left(v_0^2+u_1+u_2/v_0^2+\dots \right)^k\ ,\\ \nonumber
&&u_1=\int_{-1/2}^{1/2}dt\int_{-1/2}^{1/2}dt'\ \delta v^2(t-t')\ ,\\
&&u_2=-2\left(\int_{-1/2}^{1/2}dt\int_{-1/2}^{1/2}dt'\ \delta v^2(t-t')\right)^2+\nonumber\\&&2\int_{-1/2}^{1/2}dt\int_{-1/2}^{1/2}dt' \int_{-1/2}^{1/2}dt''\ 
\delta v^2(t-t')\delta v^2(t'-t'')\ . \nonumber
\eea
Here $u_1$ is linear in $\delta v^2$, $u_2$ is quadratic and so on. 
Expansion \eqref{rexp} became divergent at $|z|=2(v_0^2+u_1+u_2/v_0^2+\dots)$, which is the value of largest/smallest eigenvalue of $\tilde A$. Going back to \eqref{bm}, one can express $x(\Delta E)$ as
\bea
\Delta E(x)={x^2/ (8\sigma_v^2)}\ ,
\eea 
where the higher order terms in $x$ are  implicitly absorbed into a single coefficient $\sigma_v(\Delta E(x))$, 
\bea
\label{averagesigma}
\sigma_v^2=v_0^2+\int_{-1}^{1} \delta v^2(t)(1-|t|)dt+{\mathcal O}(\delta v^2)\ .
\eea
It is interesting to note that up to linear term, \eqref{averagesigma} coincides with the integral in \eqref{bs2}. Hence, when $f(E,\omega)$ is almost constant, $\sigma_v^2$ can be approximated as $\Omega \bar\Sigma^2(\Delta E)$.


\section{Canonical Universality for operators $A=i[H,B]$}
\label{sec:sY}
\begin{figure}
\includegraphics[width=.45\textwidth]{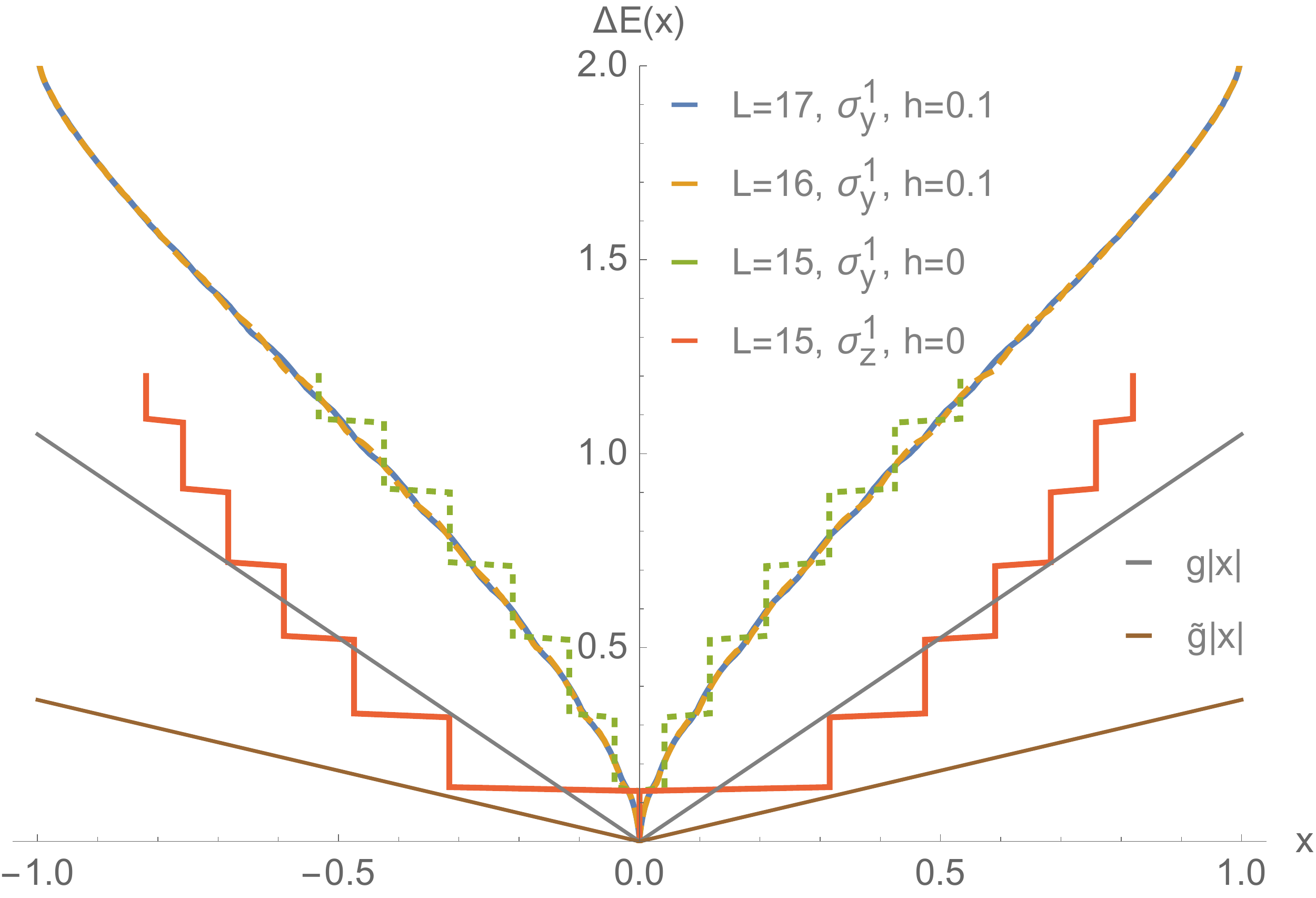}
\caption{Plot of $\Delta E(E=0,x)$ for $\sigma_y^1$ and $L=16,17$ in the non-integrable case (blue, yellow), and $L=15$ in the integrable case $h=0$ (green).  Also, plot of $\Delta E(E=0,x)$ for $\sigma_z^1$ and $L=15$ in the integrable case $h=0$ (red). Superimposed with the theoretical bound $g|x|$ (gray) and $\tilde{g}|x|$ (brown).}
\label{fig:Y}
\end{figure}

Consider an operator $A$ of the form
\bea
\label{commutator}
A=i[H, B]\ ,
\eea
for some $B$ and the Hamiltonian $H$.  If $H$ includes only local interactions and $B$ acts on a small sub-system then $A$ would be local as well. Operators of the form \eqref{commutator}, which we call  ``descendants", are special in the sense that they trivially satisfy ETH, 
\bea
\bra{E_n}A\ket{E_n}=0\ ,
\eea
which means $A^{\rm eth}(E)=0$ for any $E$.
Furthermore for any pure state $\psi$ of the form \eqref{state},
\bea
x =\bra{\psi}A \ket{\psi}= -i\bra{\psi}B\ket{\psi_1}+{\rm c.c}\ ,
\eea
where $\psi_1=(H-E)\psi$. Consequently, $|x|$ is bound from above by 
\bea
|x|\leq 2|B|\Delta E\ ,
\eea
where we  used $|\psi_1|\leq \Delta E$. This leads to the bound
\bea
\label{boundt}
\Delta E(x)\ge {|x|\over 2\,\hspace{-4pt}\parallel\hspace{-4pt}B\hspace{-4pt}\parallel}\ .
\eea
We plot $\Delta E(x)$ for $A=\sigma_y^1=i[H,\sigma_z^1/(2g)]$ together with the bound $g|x|$ in the integrable and non-integrable case in Fig.~\ref{fig:Y}. It turns out the operator $\sigma_z^1$ in the integrable case is also a descendant. The corresponding $B$ is non-local and 
\bea
(2\hspace{-4pt}\parallel\hspace{-4pt}B\hspace{-4pt}\parallel)^{-1}=:\tilde{g}(L)=g\sqrt{1-g^{-2}\over 1-g^{-2L}}\ .
\eea
Notice, that $\tilde{g}$ is finite in the infinite volume limit $L\rightarrow \infty$, and hence $\Delta E(x)$ can not be smooth at $x\rightarrow 0$. We plot $\Delta E(x)$ for $\sigma_z^1$ and the corresponding theoretical bound  in Fig.~\ref{fig:Y}.
In fact $\sigma_z^i$ for any $i$ in the integrable case is a descendant, and so is the average magnetization operator $A_z=\sum_i^L \sigma_z^i/L$. In the latter case the norm of $|B|$ grows with $L$, and therefore the bound \eqref{boundt} becomes obsolete in the thermodynamic limit. The plot for $A_z$ (Fig.~\ref{fig:SumXZ}) suggests that despite integrability $\Delta E(x)$ actually becomes a smooth function of $x$ with the characteristic $\propto x^2$ behavior at small $x$. {This is reminiscent of observation that macroscopic observables are thermal in most eigenstates for both chaotic and non-chaotic systems \cite{GoldsteinTE1,GoldsteinTE2}.}

\section{Density of states}
\begin{figure}
\includegraphics[width=.47\textwidth]{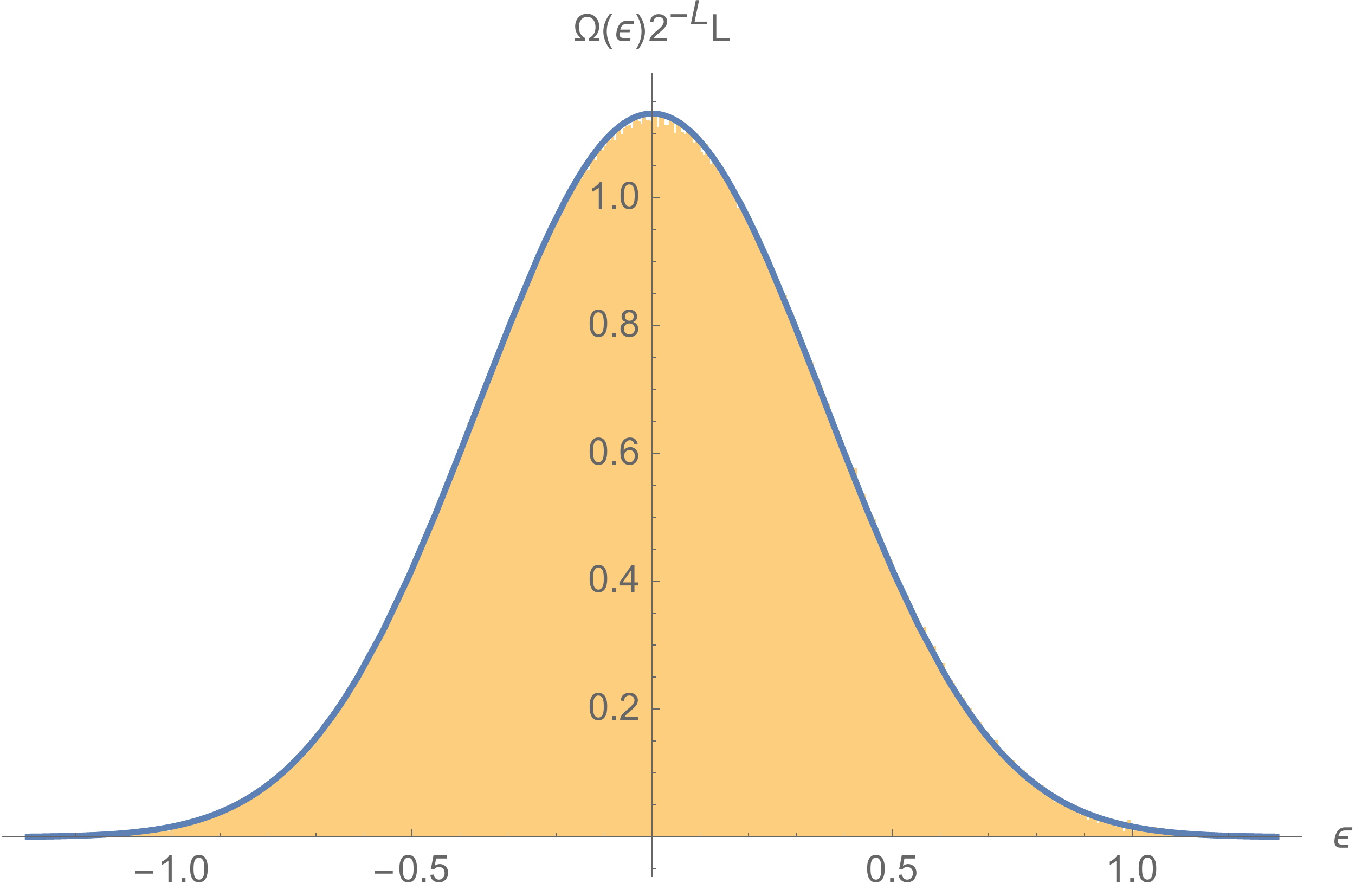}
\caption{Density of states of the spin chain \eqref{1dspinchain} with $g=1.05, h=0.1, L=17$. The horizontal axis is energy per site $\epsilon=E/L$. 
Yellow bars which is the histogram of the actual density of states calculated using direct diagonalization. The blue solid line is a theoretical 
fit by the binomial distribution function \eqref{omega} with $\kappa\approx 0.3489$. }
\label{fig:spec}
\end{figure}

The non-integrable model \eqref{1dspinchain} was numerically studied in
 \cite{Dymarskyetal}. There it was observed that the density of states is well approximated by the binomial distribution
\bea
\label{omega}
\Omega_n(E)={\kappa\, L!\over (L/2-\kappa\, E)!(L/2+\kappa\, E)!}\ ,
\eea with  $\kappa$ given by 
\bea
\label{kappa}
\kappa={1\over 2} \left(g^2+h^2+1-1/L\right)^{-1/2}\ .
\eea
The actual density of states and the theoretical fit \eqref{omega} for $L=17$ are depicted in Fig.~\ref{fig:spec}. The expression for the density of states \eqref{omega} was used to determine $\Omega^{1/2}\bar \Sigma$ shown in Fig.~\ref{fig:SO}, Fig.~\ref{fig:SX}, and Fig.~\ref{fig:SumXZS}. When $L$ becomes large $\Omega(0)$ can be approximated as 
\bea
\label{approxOmega}
\Omega(0)={2^L L^{-1/2}\over \sqrt{2\pi}\sqrt{g^2+h^2+1}}\ .
\eea
The factor $L^{-1/2}$ contributes to the correct scaling behavior of $f(E=0,\omega)$.

\section{Choice of operator $A$}
\begin{figure}
\vspace{-.6cm}
\includegraphics[width=.51\textwidth]{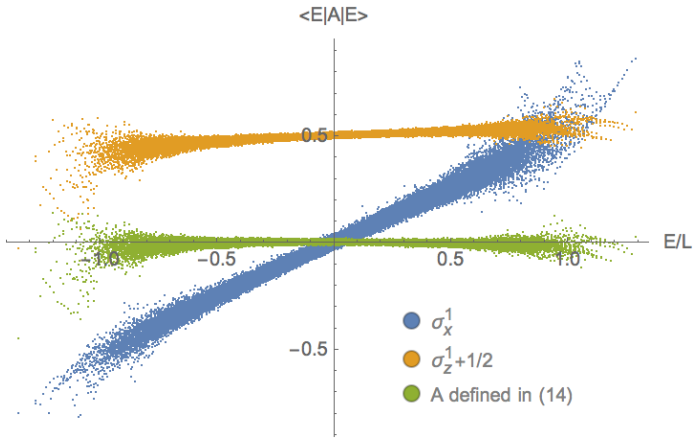}
\caption{Expectation values $\langle E_n|A|E_n\rangle$ for $A=\sigma_x^1$, $A=\sigma_z^1$, and $A$ given by \eqref{AO} in case of $L=17$, $h=0.1$ as a function of $E_n/L$.}
\label{fig:slope}
\end{figure}

For a given energy band, the value of $x(E,\Delta E)$ can not be smaller than the variations of the thermal expectation values $A^{\rm micro}(E')$ or $A^{\rm eth}(E')$ for $E'$ inside the interval $|E-E'|\leq \Delta E$. Since in the chaotic case $A^{\rm eth}(E)$ is a smooth function of $E$, these variations are of the order  $\Delta E (dA^{\rm eth}/dE)$ and  are expected to be suppressed as $1/V$ in the thermodynamic limit.  To minimize finite-size effects we would like to identify an operator with a small  value of  $dA^{\rm eth}/dE$. Looking at the one-spin operators acting on the leftmost spin (see Fig.~\ref{fig:slope}) we observe that $A^{\rm eth}(E)$ for both $\sigma_z^1$ and $\sigma_x^1$ are approximately linear function of $E$ with some non-zero slope, such that the combination \eqref{AO} has almost vanishing expectation value for a wide range of $E$ around $E=0$.

\section{Two-point function $\langle A(t) A(0)\rangle_c$} 
Function $f(E,\omega)$ can be constrained through the behavior of the connected  two-point function \cite{FDT,Review},
\bea
C(t)\equiv \langle E_n| A(t) A(0)|E_n\rangle_c=\sum_{m\neq n} e^{i(E_n-E_m)t} |A_{nm}|^2\ . \nonumber
\eea
Assuming \eqref{eth}, the integral of $C(t)$ can be rewritten as follows 
\bea
\label{mainC}
&&\int_{-\infty}^\infty dt\, C(t) {\sin(t \Delta E)\over \pi t}=\\ &&\int_{-\Delta E}^{\Delta E} d\omega  {\Omega(E_n+\omega)\over \Omega(E_n+\omega/2)}\left|f(E_n+\omega/2,\omega)\right|^2\ . \nonumber \label{wint}
\eea
In the thermodynamic limit $f(E,\omega)$ is expected to depend on $E$ only through temperature. This can be used to simplify \eqref{mainC} by neglecting $\omega$ in the first argument of  $f$. Furthermore, when $\Delta E$ is much smaller than the temperature associated with the energy $E_n$, $\omega$-dependence inside $\Omega$ also can be neglected leading to \eqref{intT}. 

When the system is substantially large $C(t)$ will smoothly depend on energy $E_n$, but not on the choice of an individual eigenvector $|E_n\rangle$. Numerically, we define $\langle  A(t) A(0)\rangle_c$ as  $\langle E_n| A(t) A(0)|E_n\rangle_c$ for $E_n=0$ by averaging over hundred states in the middle of the spectrum,
\bea
\label{defAA}
\langle  A(t) A(0)\rangle_c={1\over 100}\sum^{2^{L}/2+50}_{n=2^{L}/2-49} \langle E_n|  A(t) A(0)|E_n\rangle_c\ .\ \ 
\eea

\begin{figure}
\includegraphics[width=.475\textwidth]{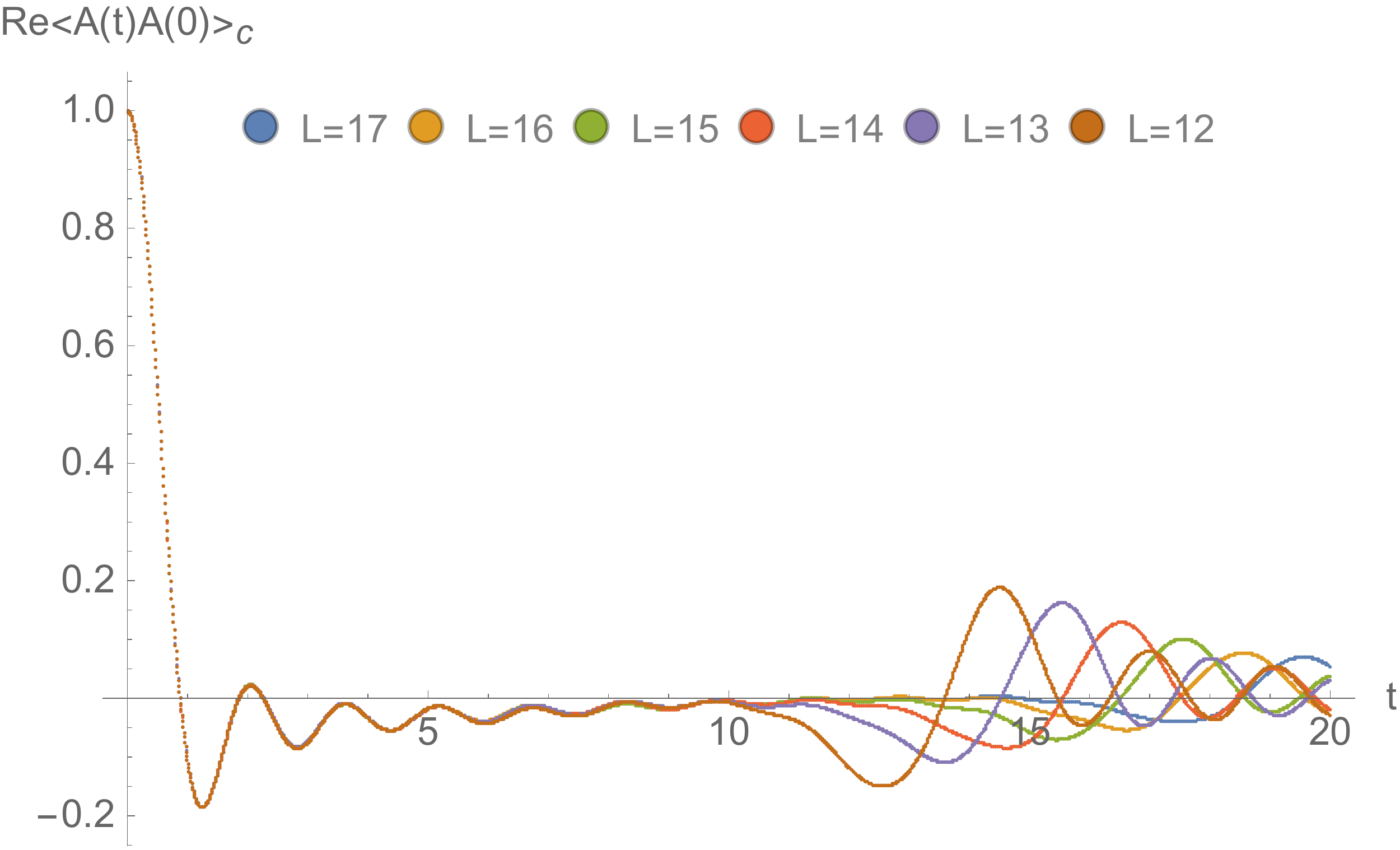}
\caption{Plot of ${\rm Re} \langle  A(t) A(0)\rangle_c$ defined in \eqref{defAA} for operator  \eqref{AO} and different $L=12-17$. } 
\label{fig:AAO}
\end{figure}
\section{Analysis of $\bar \Sigma$ and $\langle A(t) A(0)\rangle_c$ for different operators}
Here we provide additional details of the analysis of the numerical results. Based on the plot for $\bar\Sigma$  in the main text (Fig~\ref{fig:SO}) we conclude that $f(0,\omega)$ for a given $\omega$ should be $L$ independent.  The same conclusion can be reached from the analysis of two-point function $\langle A(t) A(0)\rangle_c$ shown in Fig.~\ref{fig:AAO}. In this case temperature is formally infinite, and therefore \eqref{intT}  applies so far $\Delta E\ll L^{1/2}$.
Numerical plot clearly shows that the two-point function quickly converges to an $L$-independent form for $0\leq t \leq t^*(L)$,  where  $t^*\sim L$ is the time of the ``rebound" when the finite-size effects become important. Hence for substantially large $L$, any fixed $t$ would satisfy $L^{-1/2}\ll t \ll t^*$, rendering the integral in the left-hand-side of \eqref{intT} $L$-independent. This confirms  $L$-independence of $f(0,\omega)$. 

When $\Delta E\rightarrow \infty$ the behavior of $\bar\Sigma(\Delta E)$ can be deduced from the inequality
\bea
\label{bound}
\bar\Sigma^2(\Delta E)\leq {\Tr (A^2)\over {\mathcal N}(\mathcal N-1)}\ ,
\eea
and an explicit form of $\Omega$ \eqref{approxOmega}. When $\Delta E$ is so large that the band includes almost all states, $\Omega^{1/2}(0)\bar\Sigma(0,\Delta E)$ goes to zero as $L^{-1/4}$. 

The limit of small  $\Delta E\rightarrow 0$ is more difficult to probe. For $t\ge t^*(L)$ the behavior of $\langle  A(t) A(0)\rangle_c$ is not universal, hence we can not immediately use \eqref{intT} to bound $f(0, \omega)$ in the region of small $\omega \lesssim L^{-1}$. The plot of $\Omega^{1/2}\bar\Sigma(\Delta E)$ suggests $f$ approaches a constant $f(0,\omega \rightarrow 0)= \sigma_0\sim 0.255$ (see the inset of Fig.~\ref{fig:SO}). It is nevertheless possible that in a small region of size $L^{-1}$ or less $f(E,\omega)$  grows with $L$.
\begin{figure}[t]
\includegraphics[width=.45\textwidth]{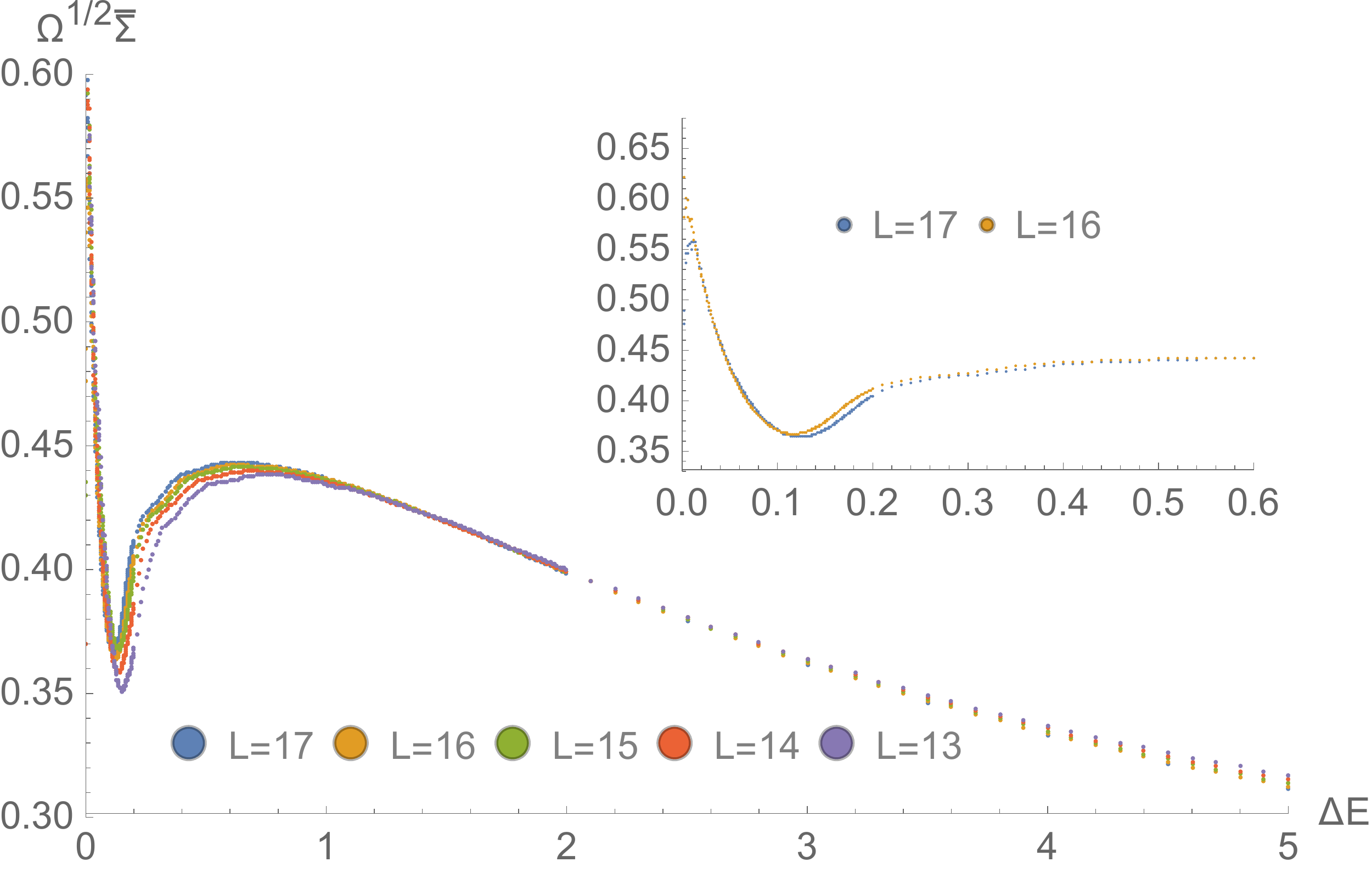}
\caption{Plot of $\Omega^{1/2}(0)\bar\Sigma(0,\Delta E)$ for $A=\sigma_x^1$, $h=0.1$ and different $L=13-17$. Inset: zoomed region of  small $\Delta E$. The limit of $\Omega^{1/2}(0)\bar\Sigma(0,\Delta E)$ as $\Delta E$ approaches zero is approximately equal to $f(0,\omega\rightarrow 0)=\sigma_0\approx 0.57$.}
\label{fig:SX}
\end{figure}

\begin{figure}[t]
\includegraphics[width=.45\textwidth]{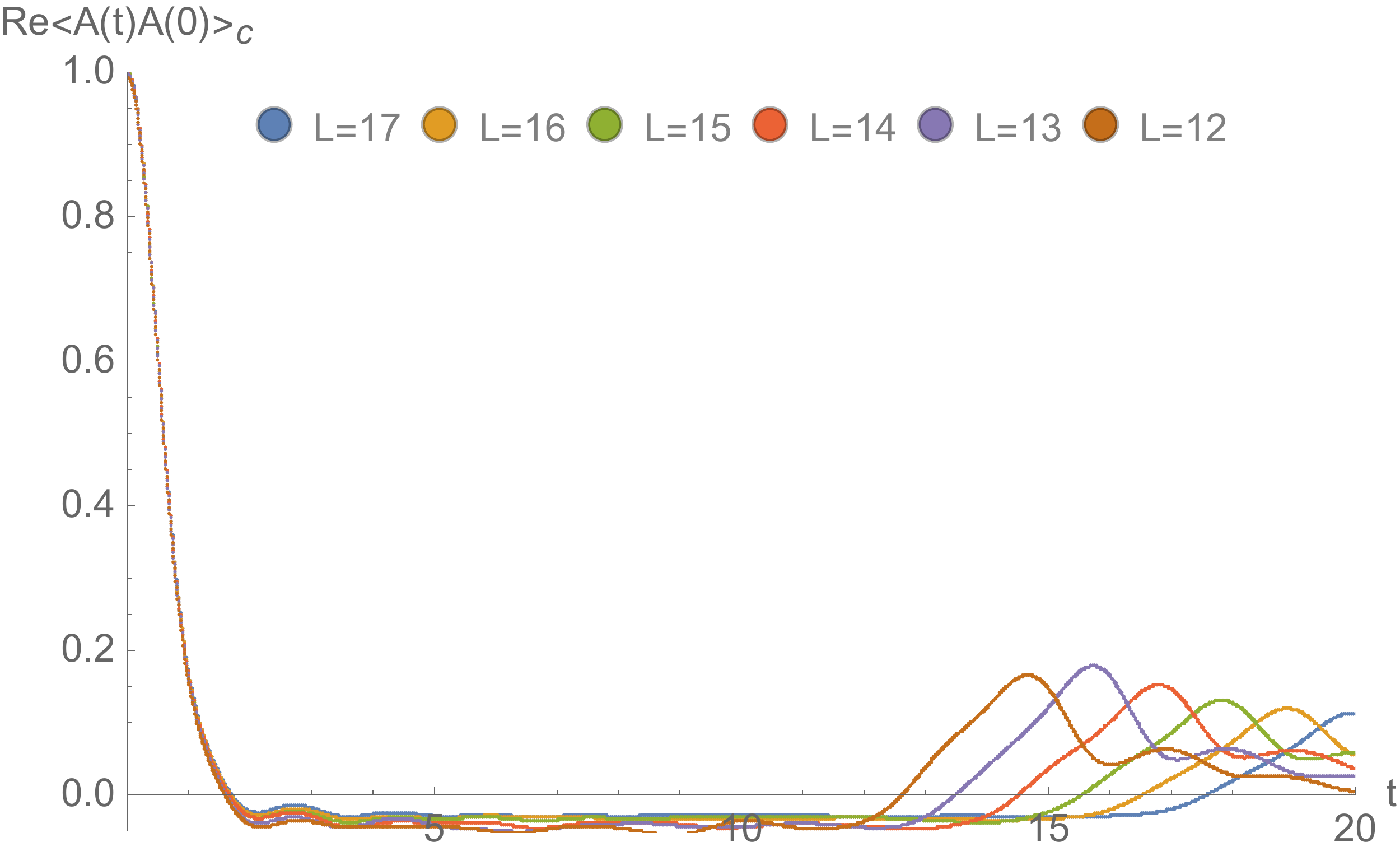}
\caption{Plot of ${\rm Re} \langle  A(t) A(0)\rangle_c$ defined in \eqref{defAA} for operator  $A=\sigma_x^1$, $h=0.1$ and different $L=12-17$.} 
\label{fig:AAX}
\end{figure}

\begin{figure}[b]
\includegraphics[width=.45\textwidth]{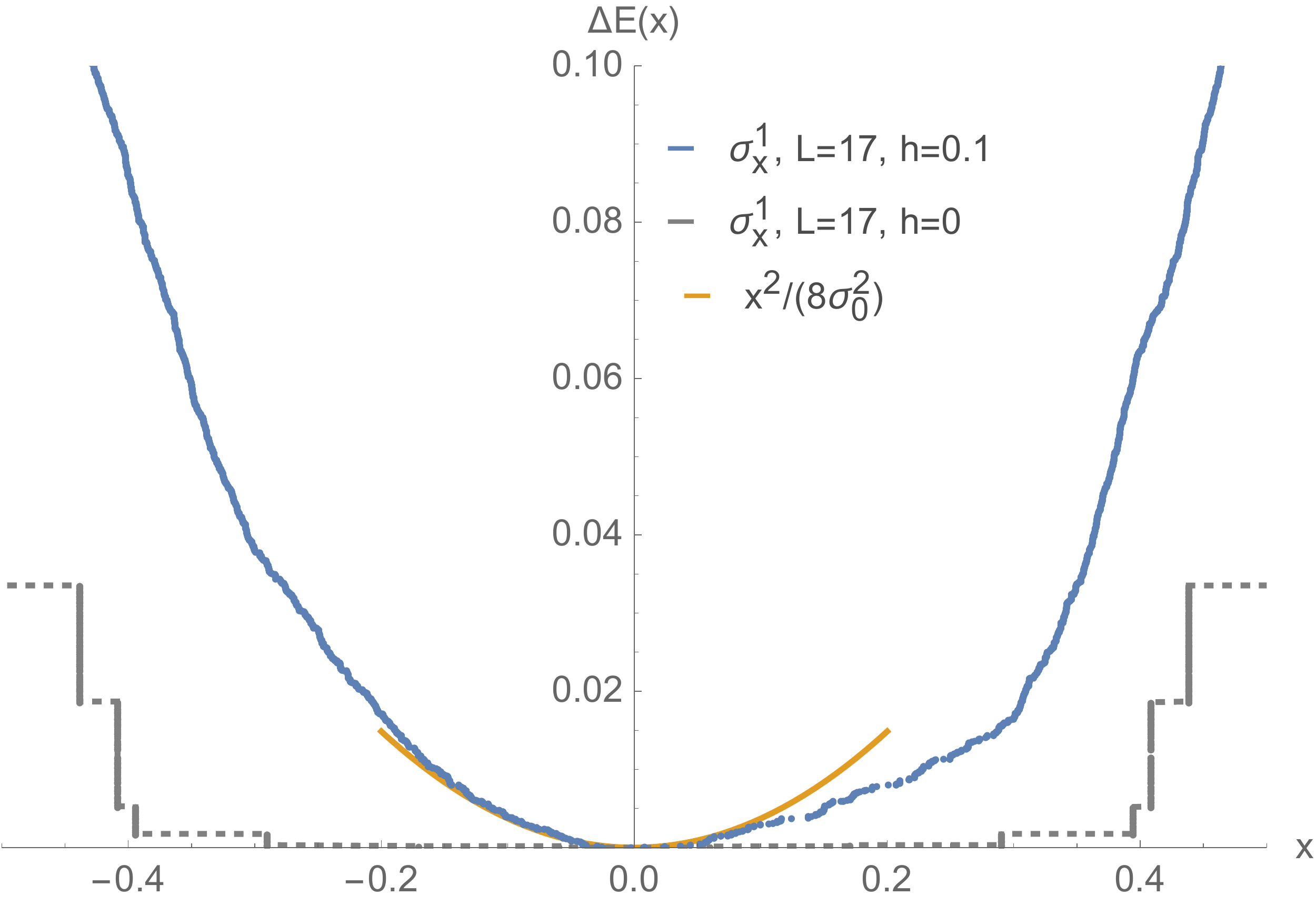}
\caption{
Plot of $\Delta E(0,x)$ for operators $\sigma_x^1$ for integrable $h=0$ (gray dashed line) and non-integrable $h=0.1$ (blue dots) cases and $L=17$, superimposed with the theoretical fit \eqref{curmm} and value of $\sigma_0\approx 0.58$ (see the inset of Fig.~\ref{fig:SX}).}
\label{fig:X}
\end{figure}

Next, we analyze one-spin operator $A=\sigma_x^1$. The corresponding plots for $\bar\Sigma$ (Fig.~\ref{fig:SX}) and $ \langle  A(t) A(0)\rangle_c$ (Fig.~\ref{fig:AAX}) support the same conclusion as above: $f(0,\omega)$ is $L$ independent in the thermodynamic limit. The plot of $\Delta E(x)$ for $\sigma_x^1$ in the integrable and non-integrable case is shown in Fig.~\ref{fig:X}. In the non-integrable case function $\Delta E(x)$ becomes smooth and is reasonably described by \eqref{curmm} at small $x$. The corresponding value of $\sigma_0\approx 0.58$ is determined as the limit of $f(0,\omega)$ as $\omega$ approaches zero, see the inset of Fig.~\ref{fig:SX}. The plot for integrable case exhibits a characteristic plateau at $\Delta E\approx 0$. Using free fermion representation of the integrable model \eqref{1dspinchain} with $h=0$, one can show the plateau at $\Delta E\approx 0$ in the thermodynamic limit $L\rightarrow \infty$ must stretch to at least $|x|\approx 0.64$. This also implies the plateau at Fig.~\ref{fig:O} will stretch to at least $|x|\approx 0.06$.

\begin{figure}[t]
\includegraphics[width=.45\textwidth]{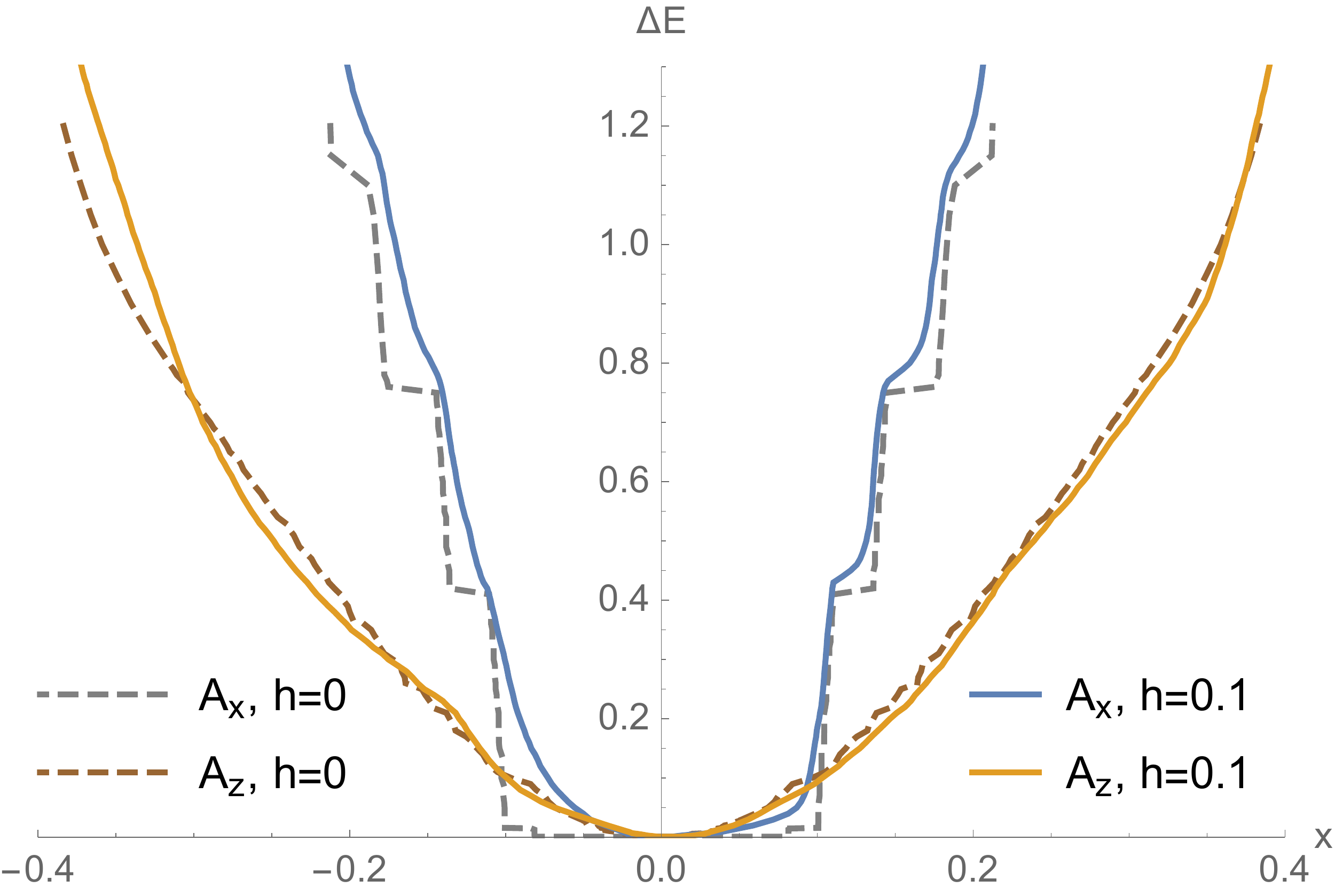}
\caption{Plot of $\Delta E(0,x)$ for operators $A_x$ and $A_z$ for integrable $h=0$ (dashed lines) and non-integrable $h=0.1$ (solid lines)  cases and $L=17$.}
\label{fig:SumXZ}
\end{figure}

As a last step we analyze extensive operators $A_{x/z}=\sum_i \sigma_{x/z}^i/L$. The plots of $\Delta E(x)$ for integrable and non-integrable cases is shown in Fig.~\ref{fig:SumXZ}. In the non-integrable case $\Delta E(x)$ for both operators is smooth and is $\propto x^2$ at small $x$. In the integrable case $\Delta E(x)$ for $A_x$ develops a characteristic plateau near $\Delta E\approx 0$ and is not smooth. The plot of $\Delta E(x)$  for  $A_z$ in the integrable case is smooth and qualitatively indistinguishable from the non-integrable case, which we assume is the consequence of $A_z$ being a descendant operator. 
The plot of $\bar\Sigma$ for $A_{x/z}$  (Fig.~\ref{fig:SumXZS}) clearly shows $L^{1/2}\Omega^{1/2}(0)\bar\Sigma(0,\Delta E)$ is $L$-independent, hence suggesting the scaling $f(0,\omega)\sim L^{-1/2}$.
 
\section{Variance of $A_{nm}$}
\label{sec:AppD}
\begin{figure}[t]
\includegraphics[width=.49\textwidth]{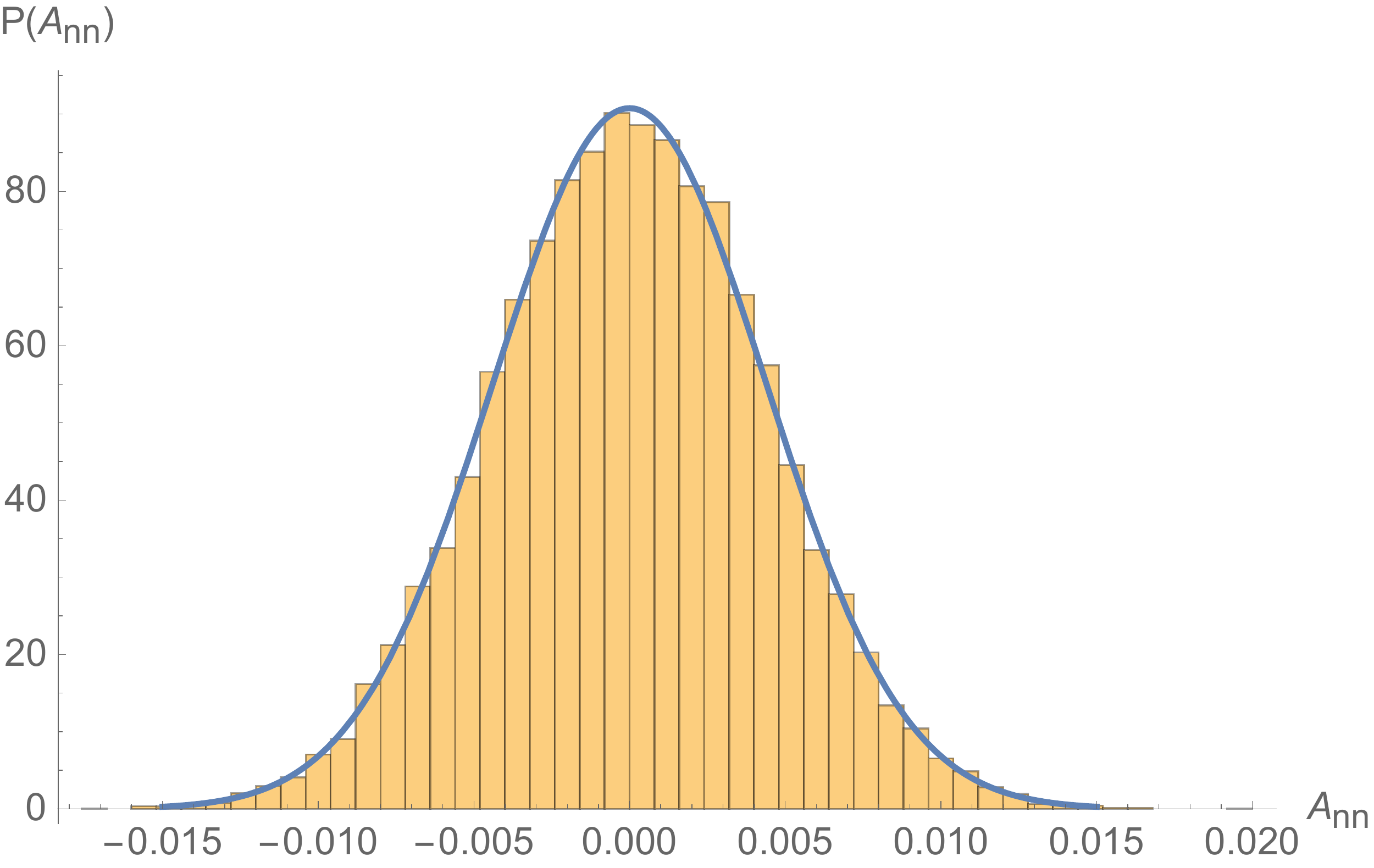}
\caption{Histogram of probability distribution of $A_{nn}$ for operator \eqref{AO} from the central band, $E=0, \Delta E= 0.1L$, for the spin-chain with $h=0.1$ of size $L=17$. Superimposed blue line is the normal distribution with the same mean and variance.}
\label{fig:histogram}
\end{figure}

\begin{figure}[t]
\includegraphics[width=.4\textwidth]{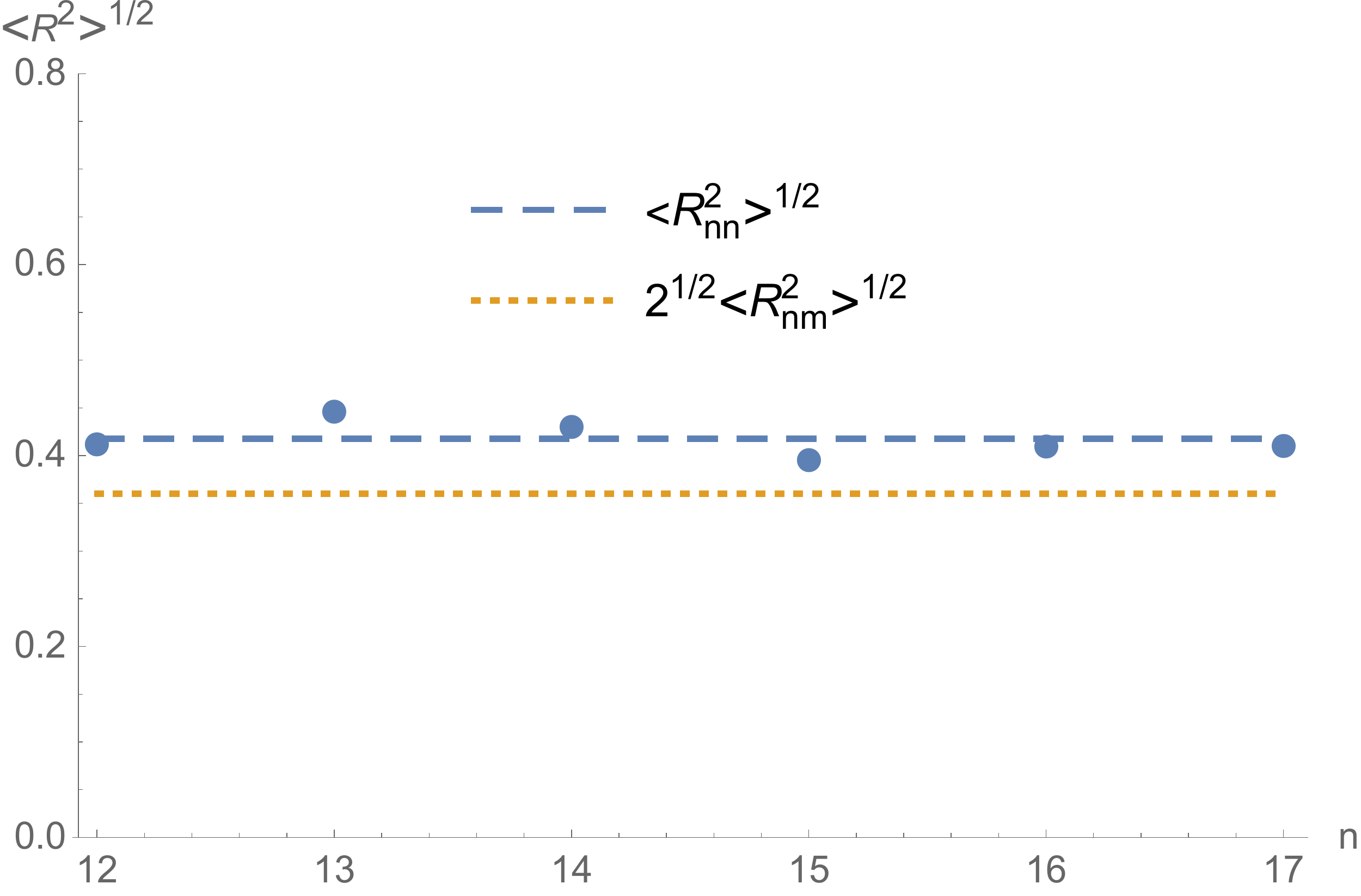}
\caption{Variance of matrix elements $R_{nn}$ for operator \eqref{AO} and $E_n$ from the central band $E=0$, $\Delta E=0.1{L}$, plotted for different values of  ${L}=12-17$. Blue dashed line: mean value $\langle R_{nn}^2\rangle^{1/2}\approx 0.418$. Dotted orange line: the  value of $2^{1/2}\langle R_{nm}^2\rangle^{1/2}=2^{1/2}\sigma_0\approx 0.361$.}
\label{fig:variance}
\end{figure}

It was observed in \cite{Dymarskyetal} that in the model in question the fluctuations of the diagonal matrix elements $A_{nn}$ of local operators are well described by the Gaussian distribution. The procedure of calculating $A^{\rm eth}(E)$  and the variance $\langle R_{nn}^2\rangle$ of $R_{nm}=\Omega^{1/2}(0)(A_{nm}-A^{\rm eth}(E_n)\delta_{nm})$ is described in \cite{Dymarskyetal}. Here we show the histogram of distribution of $R_{nn}$ inside a central band superimposed with the Gaussian fit,  see Fig.~\ref{fig:histogram}. 
The value of variance $\langle R_{nn}^2\rangle$ for \eqref{AO} and the system sizes ${ L}=12-17$  is shown in Fig.~\ref{fig:variance}. It is approximately ${L}$ independent, $\langle R_{nn}^2\rangle^{1/2}\approx 0.418$. Assuming matrix elements $R_{nm}$ inside a narrow energy band form the Orthogonal Gaussian Ensemble, variance of the off-diagonal elements, which was found in the text to be $\langle R^2_{nm}\rangle= \sigma_0^2\sim 0.255^2$ (see the inset of Fig.~\ref{fig:O}), should be twice smaller than $\langle R_{nn}^2\rangle$. This is satisfied, but only with $\sim 15\%$ accuracy: 
\bea
\label{sigma2}
\nonumber
2^{1/2}\langle R^2_{nm}\rangle^{1/2}=2^{1/2}\sigma_0\approx 0.361\ , \\
\langle R^2_{nn}\rangle^{1/2}\approx 0.418\,\ . \nonumber
\eea
This mismatch is illustrated in Fig.~\ref{fig:variance}.

\section{Calculation of $\Delta E(x)$ for a subsystem}
For a subsystem of arbitrary size and energy $E$ associated with infinite temperature, we define deviation from the microscopic thermal equilibrium by comparing reduced density matrix $\rho^\psi$ with the thermal one $\rho^{\rm th}=\mathbb{I}/d$,
\bea
x^2={1\over 2}\Tr(\rho^\psi-\rho^{\rm th})^2=(e^{-s_2(x)}-e^{-s_0})/2\ ,\\
s_2=-\log\Tr\left[ (\rho^\psi)^2\right]\ ,\quad s_0=\log d\  .
\eea
Here $d$ is the dimension of the Hilbert space of the subsystem and $s_2$ - second Renyie entropy.   To calculate $x$ as a function of state $\psi$ we introduce a full set of traceless Hermitian operators acting on the subsystem $\hat\sigma_k$, $k=1,\dots, d(d-1)$, $\Tr \hat\upvarsigma_k=0$, $\Tr (\hat{\upvarsigma}_k \hat{\upvarsigma}_\ell)=d \delta_{k\ell}$.
In case of the subsystem consisting of one spin, $d=2$ and $\hat{\upvarsigma}_k$  are simply Pauli matrices $\sigma_k$. Then $x(\psi)$ is given by 
\bea
\label{xpsi}
2d\, x^2 =\sum_{k=1}^{d(d-1)} \Tr(\rho^\psi\hat\upvarsigma_k)^2\ .
\eea
To find $x(\Delta E)$ we  need to maximize \eqref{xpsi} over all $\psi$ of the form \eqref{state}. Numerically it can be done by introducing $\mathcal N\times \mathcal N$ matrices $(\upvarsigma_k)_{nm}=\langle E_n|\hat\upvarsigma_k|E_m\rangle$, and using Lemma 2 from \cite{Convexball} to reduce the original problem to an optimization problem on a sphere, $\vec{c}\in\mathbb S^{d(d-1)}$,
\bea
\label{xdef}
x(\Delta E)={\max_{|\vec{c}|=1} \lambda_{\rm max} (\vec{c}\cdot \vec{{\upvarsigma}}) \over \sqrt{2d}}\  .
\eea
Here $\lambda_{\rm max}(\upvarsigma)$ is the largest eigenvalue of a Hermitian matrix $\upvarsigma$. In case of the leftmost spin, vector $\vec{c}\in\mathbb S^{3}$ can be conveniently parametrized with help of two angles 
\bea
\vec{c}\cdot \vec{\upvarsigma}=\cos\theta\, \sigma^1_x+\sin\theta \cos\phi\, \sigma^1_z+\sin\theta \sin\phi\,\sigma^1_y\ .
\eea
Maximization over $0\leq \theta \leq \pi$ and $0\leq \phi\leq \pi$ (it is enough to cover only half-sphere because $\lambda_{\rm max}(\upvarsigma)=\lambda_{\rm max}(-\upvarsigma)$) can be done by introducing a discretization of both intervals and then scanning through all possible values. Numerical calculations for all considered $\Delta E$ and $L$ shows that maximum of $\lambda_{\rm max}(\vec{c}\cdot \vec{\upvarsigma})$ is achieved at $\phi=0$. This is presumably related to the fact that $\sigma_y^1$  is a descendant operator and requires more energy fluctuation to deviate from thermal equilibrium. This observation substantially simplifies calculations as it reduced the problem of finding global maximum to optimization with respect to only one parameter $\theta$. The latter problem can be solved in a variety of ways, e.g.~with help of Newton's method using analytic expression for the gradient $d\lambda_{\rm max}(\vec{c}\cdot \vec{\upvarsigma})/d\theta$ in terms of eigenvectors of $\vec{c}\cdot \vec{\upvarsigma}$.

\begin{figure}[t]
\includegraphics[width=.45\textwidth]{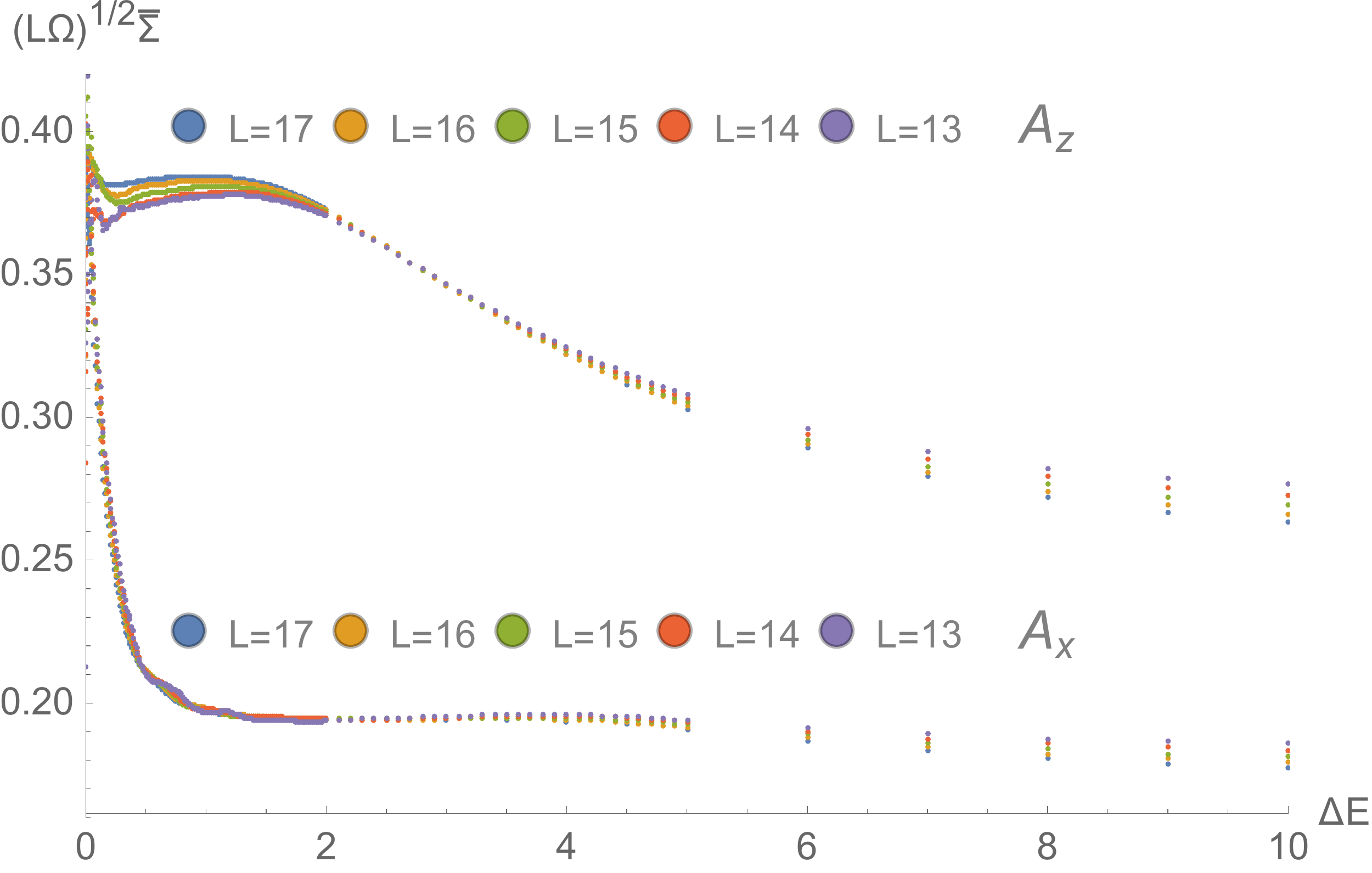}
\caption{Plot of $\Omega^{1/2}(0)\bar\Sigma(0,\Delta E)$ for operators $A_z$ (group of lines at the top) and $A_x$ (group of lines at the bottom), $h=0.1$ and different $L=13-17$.}
\label{fig:SumXZS}
\end{figure}

\section{Analytic bounds on $\Delta E(x)$}
In certain cases volume dependence of $\Delta E(x)$ can be constrained by simple analytic arguments. For example let us consider an average magnetization operator $A_{z}=\sum_i \sigma^i_{z}/L$. There is a unique state $\psi$ which maximize deviation from the thermal equilibrium, $A_{z} \psi_+=\psi_+$. State $\psi_+$ has all spins ``up" and its average energy scales with the volume $\langle \psi_+|H|\psi_+\rangle=L(h-1)+1$. Hence, $\psi_+$ would belong to an energy interval $[E-\Delta E, E+\Delta E]$ centered around $E=0$ only if $\Delta E$ scales linearly with the volume. More generally, for averaged quantities $A$, and sufficiently large deviations $x$, approaching maximal (minimal) possible values,  $\Delta E(x)/V$ should remain finite in the thermodynamic limit. This is also the behavior suggested by scaling of $|f|^2$ shown in Fig.~\ref{fig:SumXZS}.


There is another argument which bounds volume-dependence of $\Delta E(x)$ for small $x$. Let us consider a one-dimensional system \eqref{1dspinchain} or, more generally, a translationally invariant lattice model in any number of dimensions with one linear dimension $L$ taken to be much larger than all other ones. We can divide the system into two sub-systems of the respective lengths $L_1$ and $L_2$, $L=L_1+L_2$, by turning off  interacting terms in the Hamiltonian. Let us consider a state $\psi_0$ which is a tensor product of eigenstates of the corresponding subsystems
\bea
\psi_0=\ket{E_1} \otimes\ket{E_2}\ .
\eea 
From the point of view of the original system, state $\psi$ describes a state after a sudden quench when the interaction between two subsystem is turned on. Independently of values of $E_1$ and $E_2$, this state has mean energy $E=\bra{\psi_0}H\ket{\psi_0}=E_1+E_2+O(1)$, and energy variance $\delta E^2=\bra{\psi_0}(H-E)^2\ket{\psi_0}=O(1)$, where $O(1)$ indicates scaling with respect to $L$ \cite{Rigol}. 
Although $\psi_0$ may include contributions from energy eigenstates $\ket{E_n}$ with $E_n$ significantly different from $E$, an energy interval of width $\Delta E\sim \delta E$ centered around $E$ is expected to include most of the components of $\psi_0$. For the appropriate $E_1$ and $E_2$ state $\psi_0$ will bring $A$ out of equilibrium. For an averaged quantity $A$, in the limit  $L_1,L_2\rightarrow \infty$ expectation value $\bra{\psi_0}A\ket{\psi_0}$ is given by 
\bea
\label{psiev}
\bra{\psi_0}A\ket{\psi_0} = {L_1 A^{\rm eth}(E_1/L_1)+L_2 A^{\rm eth}(E_2/L_2)\over L}\ .
\eea
The deviation from the thermal equilibrium $x$ is 
the difference between \eqref{psiev} and thermal expectation value $A^{\rm eth}((E_1+E_2)/L)$. Taking thermodynamic limit while keeping the energy density $E_i/L_i=\epsilon_i$ and ratio $\lambda=L_1/L$ fixed, we find 
\bea
\nonumber
&&x=\lambda A^{\rm eth}(\epsilon_1)+(1-\lambda) A^{\rm eth}(\epsilon_2)- A^{\rm eth}(\lambda \epsilon_1+(1-\lambda)\epsilon_2)\ ,\\
&&E/L=\lambda \epsilon_1+(1-\lambda) \epsilon_2\ .
\label{eng}
\eea
We see that deviation $x$ measures deviation of $A^{\rm eth}(\epsilon)$ from a straight line. In general $x$ is finite in the thermodynamic limit. This argument shows that for $x$ small enough, such that it can be achieved for a given $E$ through \eqref{eng} by choosing an appropriate $\epsilon_1,\epsilon_2,\lambda$, necessary interval width $\Delta E(E,x)$ will not grow with $L$.

The combination of these two arguments readily shows that for a typical averaged quantity $A$, $\Delta E(x)$ should scale differently with  $V$ for different values of $x$. 
We expect different scaling of $\Delta E(x)$ for different $x$ also to apply for local operators as well. 


\begin{thebibliography}{99}



\bibitem{Goldstein}
S.~Goldstein,  J.~Lebowitz, R.~Tumulka, and N.~Zanghi, ``Canonical typicality,'' Physical review letters 96, no. 5 (2006): 050403, [arXiv:cond-mat/0511091].

\bibitem{Popescu}
S.~Popescu, A.~Short , A.~Winter, ``Entanglement and the foundations of statistical mechanics,'' Nature Physics, (2006): 2(11), 754-758, [arXiv:quant-ph/0511225].


\bibitem{Deutsch}
J.~Deutsch, ``Quantum statistical mechanics in a closed system,'' Physical Review A 43, no. 4 (1991): 2046.


\bibitem{Srednicki}
Srednicki, ``Chaos and quantum thermalization,'' Physical Review E 50, no. 2 (1994): 888.

\bibitem{Srednicki1999}
M.~Srednicki, ``The approach to thermal equilibrium in quantized chaotic systems," Journal of Physics A: Mathematical and General 32.7 (1999): 1163.



\bibitem{Rigol}
M.~Rigol, V.~Dunjko, and M.~Olshanii,  ``Thermalization and its mechanism for generic isolated quantum systems,'' Nature,  (2008): 452(7189), 854-858, [arXiv:0708.1324].



\bibitem{GoldsteinTE1}
S.~Goldstein, D.~Huse, J.~Lebowitz, R.~Tumulka, ``Thermal equilibrium of a macroscopic quantum system in a pure state," Physical Review Letters. 2015 Sep. 4;115(10):100402, [arXiv:1506.07494].

\bibitem{GoldsteinTE2}
S.~Goldstein, D.~Huse, J.~Lebowitz, R.~Tumulka, ``Macroscopic and Microscopic Thermal Equilibrium," [arXiv:1610.02312]


\bibitem{SETH}
H.~Kim, T.~Ikeda, D.~Huse, ``Testing whether all eigenstates obey the Eigenstate Thermalization Hypothesis," Phys. Rev. E 90, 052105 (2014), [arXiv:1408.0535].

\bibitem{Dymarskyetal}
A.~Dymarsky, N.~Lashkari, H.~Liu, ``Subsystem ETH", [arXiv:1611.08764].

\bibitem{FDT}
E.~Khatami, G.~Pupillo, M.~Srednicki, M.~Rigol,
``Fluctuation-Dissipation Theorem in an Isolated System of Quantum Dipolar Bosons after a Quench," Phys. Rev. Lett. 111, 050403 (2013), [arXiv:1304.7279]. 

\bibitem{Review}
L.~D'Alessio, Y.~Kafri, A.~Polkovnikov, M.~Rigol,
``From Quantum Chaos and Eigenstate Thermalization to Statistical Mechanics and Thermodynamics," Adv. Phys. 65, 239 (2016), [arXiv:1509.06411].


\bibitem{diagonal}
W.~Beugeling, R.~Moessner, and M.~Haque, ``Finite-size scaling of eigenstate thermalization," Phys. Rev. E 89, 042112 (2014), [arXiv:1308.2862].

\bibitem{offdiagonal}
W.~Beugeling, R.~Moessner, M.~Haque,
``Off-diagonal matrix elements of local operators in many-body quantum systems,"
Phys. Rev. E 91, 012144 (2015), [arXiv:1407.2043].

\bibitem{Pastur}
S.~Molchanov, L.~Pastur, and A. Khorunzhii, ``Limiting eigenvalue distribution for band random matrices," Theoretical and Mathematical Physics 90, no. 2 (1992): 108-118.

\bibitem{Convexball}
 A.~Dymarsky, ``Convexity of a Small Ball Under Quadratic Map," Linear Algebra and Its Applications, Volume 488, (2016), p. 109–123, [arXiv:1410.1553].

\bibitem{gamma}
D.~Luitz, Y.~Bar Lev, ``Anomalous thermalization in ergodic systems," Phys. Rev. Lett. 117, 170404 (2016), [arXiv:1607.01012].



\end{thebibliography}
\end{document}